\documentclass[aps,prl,reprint,superscriptaddress]{revtex4-1}

\usepackage{graphicx}% Include figure files
\usepackage{epstopdf}

\begin{document}

\title{Two Superconducting Transitions in Single Crystal La$_{2-x}$Ba$_x$CuO$_4$}

\author{X. Y. Tee}
\affiliation{Division of Physics and Applied Physics, School of Physical and Mathematical Sciences, Nanyang Technological University, 637371 Singapore.}

\author{T. Ito}
\affiliation{National Institute of Advanced Industrial Science and Technology, Tsukuba, Ibaraki 305-8562, Japan.}

\author{T. Ushiyama}
\affiliation{National Institute of Advanced Industrial Science and Technology, Tsukuba, Ibaraki 305-8562, Japan.}

\author{Y. Tomioka}
\affiliation{National Institute of Advanced Industrial Science and Technology, Tsukuba, Ibaraki 305-8562, Japan.}

\author{I. Martin}
\affiliation{Materials Science Division, Argonne National Laboratory, Argonne, Illinois 60439, USA.}

\author{C. Panagopoulos}
\affiliation{Division of Physics and Applied Physics, School of Physical and Mathematical Sciences, Nanyang Technological University, 637371 Singapore.}
\affiliation{Department of Physics, University of Crete and FORTH, GR-71003 Heraklion, Greece}

\date{\today}

\begin{abstract}
We use spatially-resolved transport techniques to investigate the superconducting properties of single crystals La$_{2-x}$Ba$_x$CuO$_4$.
We find a new superconducting transition temperature $T_{cs}$ associated with the ab-plane surface region which is considerably higher than the bulk $T_c$.
The effect is pronounced in the region of charge carrier doping $x$ with strong spin-charge stripe correlations, reaching $T_{cs}=36$ K or 1.64$T_c$.

\end{abstract}

%\keywords{}

\maketitle
The first high-$T_c$ superconductor La$_{2-x}$Ba$_x$CuO$_4$ discovered 30 years ago is still a subject of active research.
It is believed that clarifying the interplay between the superconducting and the charge and spin orders in this material can provide key insights into the mechanism of high-$T_c$ superconductivity.
However, unlike its sister compound YBa$_2$Cu$_3$O$_{7-x}$, this material is still not as well characterized due to the historic lack of high quality single crystals.
Recent advances in crystal growth techniques however, enabled detailed investigations on high quality samples.
Here we show, by carefully isolating the transport properties of the ab-plane surface region from the sample bulk, that the surface region undergoes superconducting transition at $T_{cs}$ which is considerably higher than the bulk $T_c$. 
This effect occurs in the range of doping where the system is known to have the strongest tendency to form charge/spin stripes \cite{Tranquada95,Abbamonte05,Fujita04,Dunsiger08,Hucker11}{}. 
Notably, the largest enhancement - 64$\%$ - we observed was at x=0.12, where the bulk stripe correlations are presumed to be most pronounced.

High quality single crystals of La$_{2-x}$Ba$_x$CuO$_4$ (LBCO-$x$) over the composition range $0.076<x<0.139$ were grown with the newly developed laser-diode-heated floating zone (LDFZ) method which enabled exceptionally high degree of homogeneity \cite{Ito13}. 
Laue and x-ray diffraction indicated single crystallinity and absence of impurity phases. 
The samples were cut into bar shape of typical dimensions 3.0 mm x 0.5 mm x 0.5 mm, with the last dimension along the crystal c-axis.  
Prior to the deposition of silver paste for contact formation, the samples were polished with diamond lapping film of grit size 1 $\mu$m followed by thorough cleaning with ethanol at room temperature.
Such polishing can cause surface roughness on a micron scale; note, however, that a continuous superconducting path can be established even when superconducting layer thickness $d$ is much smaller than surface roughness, due to the c-axis Josephson coupling. From our analysis we find that $d$ can be as small as nanometer.
Distinction between crystal a- and b-axes was omitted as the samples were not detwinned. 
These materials have a layered structure with copper-oxide planes stacked along the c-axis. 
In this work, the in-plane transport properties near the ab-plane surface and in the bulk are probed and differentiated with voltage contacts placed on the crystal ab or ac faces, respectively (Supplementary Figs. S1-2). 
In either configuration, the voltage contacts were kept a distance $\sim\mu$m from the sample edges to avoid pick-up of the counterpart signals, while electrical or heat current was applied through the entire bc faces.
In the rest of this paper, the subscript $s$ and $ab$ are used to denote surface and bulk properties, respectively.

Figures 1a-e show the temperature dependence of surface ($\rho_{s}$) and bulk ($\rho_{ab}$) resistivity for various $x$. For LBCO-0.076 and LBCO-0.092, $\rho_{s}$ and $\rho_{ab}$ drop to zero at $T_{c}$, as expected for a bulk superconductor. 
For LBCO-0.115, LBCO-0.120, and LBCO-0.139, a kink is observed in both $\rho_{s}$ and $\rho_{ab}$ at the low-temperature-tetragonal (LTT) structural transition temperature $T_{d}$\cite{Hucker11}{}. 
The striking observation here is that $\rho_{s}$ and $\rho_{ab}$ drop to zero at two different critical temperatures namely, $T_{cs}$ and $T_{c}$, with the former significantly higher than the latter. In particular, for LBCO-0.120, $T_{cs}=36$ K which is 64 $\%$ higher than $T_{c}=22$ K. 
Figure 1f shows the temperature dependence of surface ($S_{s}$) and bulk ($S_{ab}$) thermopower for LBCO-0.120. 
At high temperatures, the two curves overlap with a rapid drop at $T_{d}$. As the temperature decreases further, $S_{s}$ drops to zero at $T_{cs}=36$ K, whereas $S_{ab}$ crosses from positive to negative before it eventually becomes zero at $T_{c}=22$ K. The zero-crossing of $S_{ab}$ has been recently attributed to electronic reconstruction which produces charge carriers of different signs at low temperatures \cite{Taillefer10,Taillefer07}{}. 
Both electrical resistivity and thermopower consistently point to a higher $T_{cs}$ near the surface. The effect was observed on the opposite surface of the crystal. Furthermore, it is reproducible in different samples of the same doping (Fig. S3).

Magneto-resistivity measurements on LBCO-0.120 with magnetic fields ($H$) applied along the c-axis reveal that superconductivity near the surface is more robust than in the bulk (Fig. 2a). Upon increasing $H$, both $T_{cs}$ and $T_{c}$ decrease; however, $T_{cs}/T_{c}$ is found to increase.
At the highest applied magnetic field ($H=14$ T), both $\rho_{s}$ and $\rho_{ab}$ show insulating behaviour ($d\rho/dT<0$) before the onset of superconductivity, characteristic of underdoped lanthanum-based cuprates \cite{Ando95}{}. 
The variation of $T_{cs}$ and $T_{c}$ with magnetic field is consistently observed in magneto-thermopower measurements (Fig. 2b).
We also measured Nernst effect which is a sensitive probe of superconducting vortex flow (Supplementary Figs. S4-6). In the measurement, a magnetic field is applied along the c-axis.
The flow of vortices driven by a temperature gradient generates an electric signal along the traverse direction. 
For LBCO-0.120, the surface ($\nu_{s}$) and bulk ($\nu_{ab}$) Nernst coefficients peak around $T_{cs}$ and $T_{c}$, respectively, consistent with the vortex flow mechanism of the Nernts effect \cite{Wang06}. 
Below the transition temperatures, $\nu_{s}$ and $\nu_{ab}$ are suppressed and eventually vanish due to the pinning of vortices. 
The observation of vortex pinning below $T_{cs}$ provides evidence of phase-coherent superconductivity near the surface.

We verified sample homogeneity by comparing the normal state properties of the surface region and the bulk.
Thermopower (Fig. 1f) and Nernst coefficient (Supplementary Fig. S5) of the two regions overlap prior to the onsets of superconductivity. 
When superconductivity is suppressed and normal state recovered by strong magnetic fields, the overlap of surface and bulk thermopower persists to even lower temperatures (Fig. 2b).
For resistivity measurements, after unavoidable pickup of $\rho_c$ taken into account, excellent scaling is observed between $\rho_s$ and $\rho_{ab}$ at temperatures higher than $T_{cs}$ (Supplementary Fig. S7). The scaling holds also in resistance measured across different parts of the sample (Supplementary Figs. S8-9)

Surface I-V characteristics and $\rho_{s}$ are well fitted by assuming the Berezinskii-Kosterlitz-Thouless (BKT) transition, which describes 2D superfluid systems (Supplementary Fig. S10). 
This indicates the higher $T_{cs}$ superconductivity is either confined to a thickness not exceeding the superconducting coherence length $\xi$ (comparable to the atomic lattice constant of this material), or originates from a set of superconducting planes with no mutual phase coherence but an overall thickness larger than $\xi$ \cite{Li07}. 
While we lack the tools for direct determination of the thickness of the superconducting surface layer, based on our analysis of the surface critical current, the superconducting surface region can be as thin as a few nanometers (Supplementary Fig. S11).

Model calculation shows the superconducting surface region does not short the resistive bulk, due to the large c-axis and finite current-contact resistances (Supplementary Fig. S12-14). 
This is consistent with our observation that the bulk remains resistive even while being ``sandwiched'' by the superconducting surfaces over the temperature range $T_{c}<T<T_{cs}$, and becomes superconducting only below $T_{c}$. 
When voltage contacts cover simultaneously the crystal ac and ab faces, a contribution from \emph{both} the surface region and the bulk was detected. 
For instance, the measured resistance showed a partial drop at $T_{cs}$ prior to the full superconducting transition at $T_{c}$ (Supplementary Fig. S15).

We now discuss possible origins of this striking  behavior.
The two distinct superconducting transitions could in principle result from chemical inhomogeneity.
For instance, the surface region may have an effective concentration of Ba ions, $x_{s}$, which is different from its bulk counterpart $x$.
Since bulk $T_{c}$ correlates with $x$, this could lead to the higher $T_{cs}$. 
However, earlier investigation on the bulk superconductivity of LBCO revealed that the maximum $T_{c}^{max}=32$ K is reached for $x=0.095$ \cite{Hucker11}{}. 
Thus, the fact that $T_{cs}$ of LBCO-0.115, LBCO-0.120, and LBCO-0.139 are higher than bulk $T_{c}^{max}$ cannot be explained by an effective $x_{s}$.
We also note that polishing the surface had no effect: the superconducting surface region was reproducibly observed in a sample after several surface re-polishing over a period of approximately two years.
If the enhanced superconductivity is due to a surface region of different chemical composition, the polishing would have removed it - which is not the case here.

The phase diagram of LBCO (Fig. 3) is known to show a strong suppression of bulk superconductivity over the composition range $0.095<x<0.155$, with the strongest effect at $x=1/8$ where $T_{c}$ is approximately 5 K \cite{Hucker11,Moodenbaugh88}{}. 
This effect is known as the ``1/8 anomaly'' and has been attributed to the stabilization of stripe order, which is the segregation of charge carriers (holes) into hole-rich unidirectional charge stripes, forming antiphase boundaries
between hole-poor antiferromagnetically ordered spin domains \cite{Tranquada95,Abbamonte05,Fujita04,Dunsiger08,Hucker11}{}.
Surprisingly, this is the region where we observed the higher $T_{cs}$ for x = 0.115, 0.12 and 0.139. 
For x = 0.076 and 0.092, where the stripes are fluctuating \cite{Fujita04,Dunsiger08}, the effect was not observed.

The coincidence of the observed effect with stripe order suggests two likely possibilities.
Since surface represents an abrupt structural termination, it is possible that the crystal structure near the surface differs significantly from the bulk \cite{Oura03}. 
It is known that suppression of the LTT phase by strain enhances $T_{c}$ of LBCO thin films \cite{Sato00b}. It is possible that the surface region does not transform into the LTT structure and hence has a higher $T_{cs}$. 
One reason why LTT structure may be detrimental to superconductivity is that it is associated with stripe stabilization \cite{Fujita04}: When carriers are tied up in static stripes they cannot effectively participate in superconductivity. 
Without the stabilization effect of the LTT structure, stripes near the surface may become dynamic \cite{Fujita04} in which case, as it has been suggested \cite{Kivelson98}, superconductivity may be enhanced as well.
In addition to the structural change scenario, one may also speculate that the electronic structure near the surface may be different from the bulk \cite{Oura03}.
This may happen, for example, due to different degree of stripe stabilization in the two regions. 
If the carrier density is higher near the surface, then proximity coupling to the underdoped bulk could lead to the enhanced values of $T_{cs}$ \cite{Kivelson02,Yuli08}{}.

The higher $T_{cs}$ superconductivity was not identified in earlier transport measurements of LBCO \cite{Li07,Adachi01,Adachi05,Adachi11}{}. 
There are, however, significant discrepancies among published results. 
A partial drop in $\rho_{ab}$ at a temperature above $T_{c}$ was reported for $x=1/8$ and attributed to the onset of superconducting fluctuations \cite{Li07}{}.
However, this behaviour was absent in other works \cite{Adachi01,Adachi05,Adachi11}{}.
This discrepancy may be due to the difference in the contact configurations. 
For example, the partial drop in resistivity in the paper of Li et al. \cite{Li07} was possibly due to the voltage contacts covering the surface region in the measurement (Supplementary Fig. S15). 
We find that confining the voltage contacts to the side of the sample removes the partial drop, revealing the actual $\rho_{ab}$.
Our results suggest re-interpretation of the previous transport experiments on LBCO. 
It is the isolation of the ab-plane surface and bulk regions by different contact configurations that enabled us to differentiate the surface and bulk electronic transport channels and to detect $T_{cs}$.

Our findings may provide  an insight into the reported spectroscopic data taken on the ab-plane surface of LBCO-0.125.
Angle-resolved photoemission (ARPES) and scanning tunneling spectroscopy (STS) experiments indicate the opening of a $d$-wave gap $\Delta_0$ prior to the bulk superconducting transition \cite{Valla06,He09,Valla12}.
We note, this energy gap correlates well with the higher $T_{cs}$ superconductivity since (a) the latter takes place near the ab-plane surface, (b) the amplitudes of the two effects peak around $x\approx 1/8$ and show similar trend in the $T-x$ phase diagram, and (c) the onset temperatures of $\Delta_0$ and $T_{cs}$ coincide, at least for $x\approx 1/8$ which is around 40 K.
This remarkable consistency suggests the spectroscopic $d$-wave gap could be due to the higher $T_{cs}$ surface superconductivity.

In summary, we have presented the  experimental observation of a ab-plane surface superconducting transition $T_{cs}$ which is considerably higher than the known bulk $T_c$ in single crystal LBCO. 
Across the range of charge carrier doping concentration studied in this work, the effect is most prominent when multiple correlated electronic/structural orders are simultaneously present, pointing at strong correlations as the key ingredient. 
While the precise mechanism of our observations remains an open problem, our findings suggest that crystal surfaces can be the locus of nontrivial interplay of superconducting and non-superconducting orders.  
Indeed, similar, albeit less dramatic, intermediate doping suppression of superconductivity is observed in several other cuprates, including La$_{2-x}$Sr$_x$CuO$_4$ \cite{Drachuck12}, La$_{2-x-y}$Nd$_y$Sr$_x$CuO$_4$ \cite{Nakamura92}, and YBa$_2$Cu$_3$O$_{6+x}$ \cite{Liang06}.
For YBa$_2$Cu$_3$O$_{6+x}$, there has been indication that bulk $T_c$ can be enhanced by applied pressure \cite{Cyr15}.
These and other correlated superconductors are thus promising candidates for observation of a higher $T_{cs}$.

\begin{acknowledgments}
This work was supported by the National Research Foundation, Singapore, through a Fellowship and Grant NRF-CRP4-2008-04. The work at AIST was supported by JSPS Grants-in-Aid for Scientific Research (Grant no. 22560018). The work at Argonne was supported by U.S. Department of Energy, Office of Science, Materials Sciences and Engineering Division. The work in Greece was financed by the European Union (European Social Fund, ESF) and Greek national funds through the Operational Programme “Education and Lifelong Learning” of the National Strategic Reference Framework (NSRF) under “Funding of proposals that have received a positive evaluation in the 3rd and 4th Call of ERC Grant Schemes”. We thank A. P. Petrovi\'{c} and A. Soumyanarayanan for helpful discussions and X. Xu for assistance with the experimental set up.
\end{acknowledgments}

\begin{figure*} [h]
\includegraphics [scale=0.8] {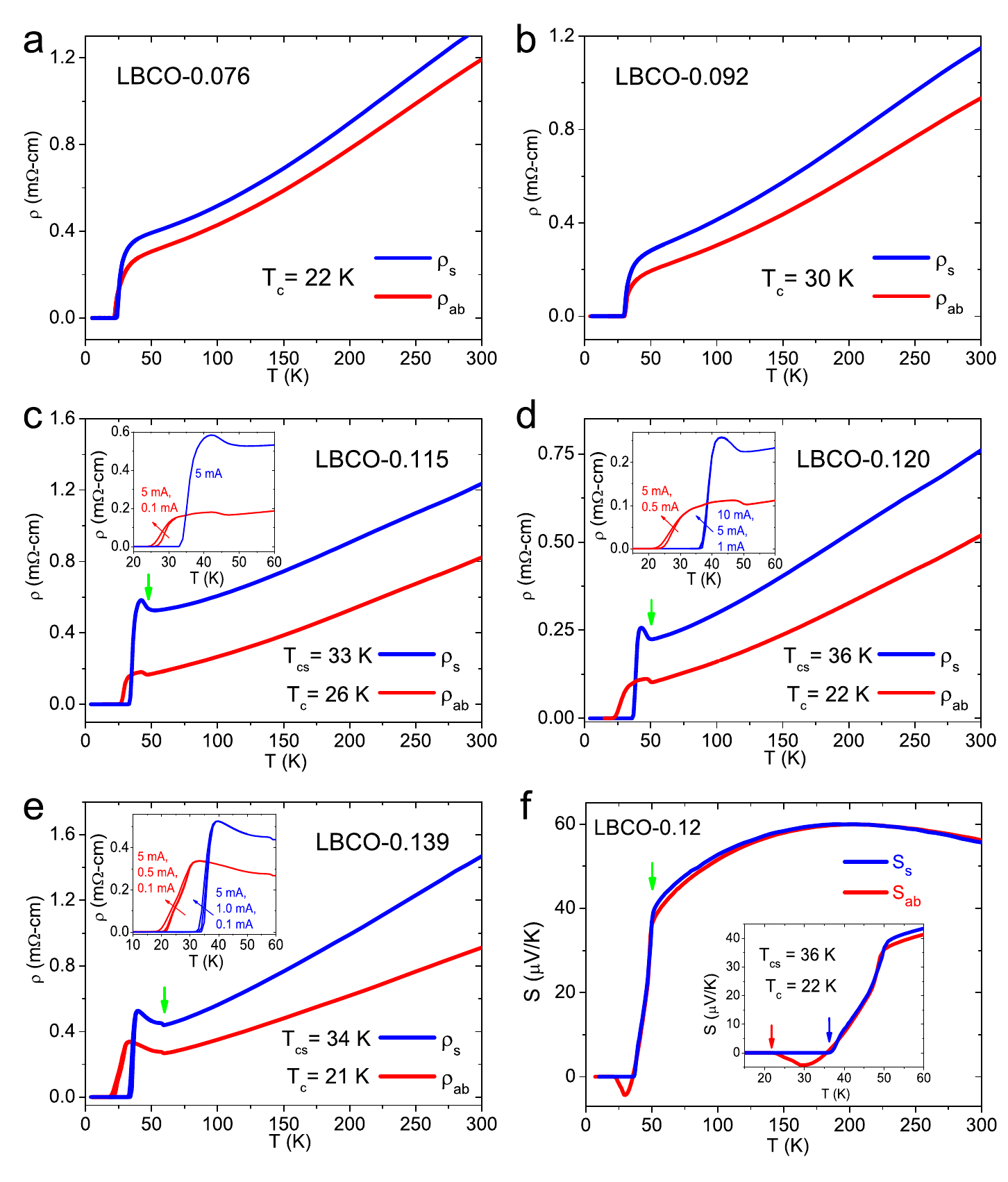}
\caption{(a)-(e) Temperature dependence of $\rho_{s}$ and $\rho_{ab}$ for various compositions. For LBCO-0.076 and LBCO-0.092, $\rho_{s}$ and $\rho_{ab}$ drop to zero at a unique $T_c$. For LBCO-0.115, LBCO-0.120, and LBCO-0.139, $\rho_{s}$ and $\rho_{ab}$ drop to zero at $T_{cs}$ and $T_c$, respectively. The green arrows in \textbf{c}-\textbf{e} indicate the structural transition at $T_d$. Insets of \textbf{c}-\textbf{e} show magnified views of $\rho_{s}$ and $\rho_{ab}$ measured with different currents near the superconducting transitions. (f) Temperature dependence of $S_{s}$ and $S_{ab}$ for LBCO-0.120. $S_{s}$ and $S_{ab}$ drop to zero at $T_{cs}=36$ K and $T_c=22$ K respectively, consistent with the results of resistivity measurement. Inset shows magnified view of $S_{s}$ and $S_{ab}$ near the superconducting transitions indicated by blue and red arrows, respectively. }
\end{figure*}

\begin{figure*}
\includegraphics [scale=0.5] {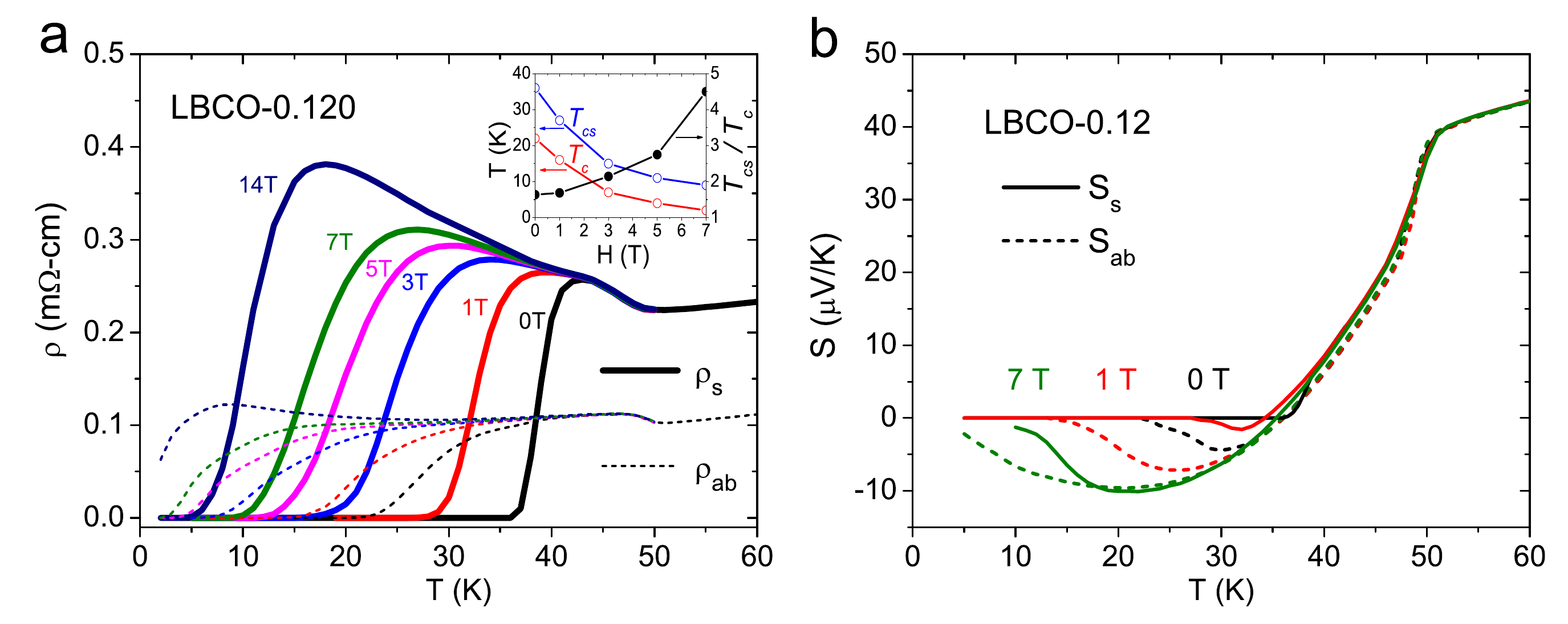}
\caption{(a) Temperature dependence of $\rho_{s}$ and $\rho_{ab}$ measured in various magnetic fields along the c-axis. Inset shows $T_{cs}$, $T_{c}$, and $T_{cs}/T_{c}$ as a function of magnetic field, indicating that superconductivity near the surface is more robust with respect to applied magnetic field than in the bulk. (b) Temperature dependence of $S_{s}$ and $S_{ab}$ measured in various magnetic fields along the c-axis.}
\end{figure*} 

\begin{figure*}
\includegraphics [scale=0.35] {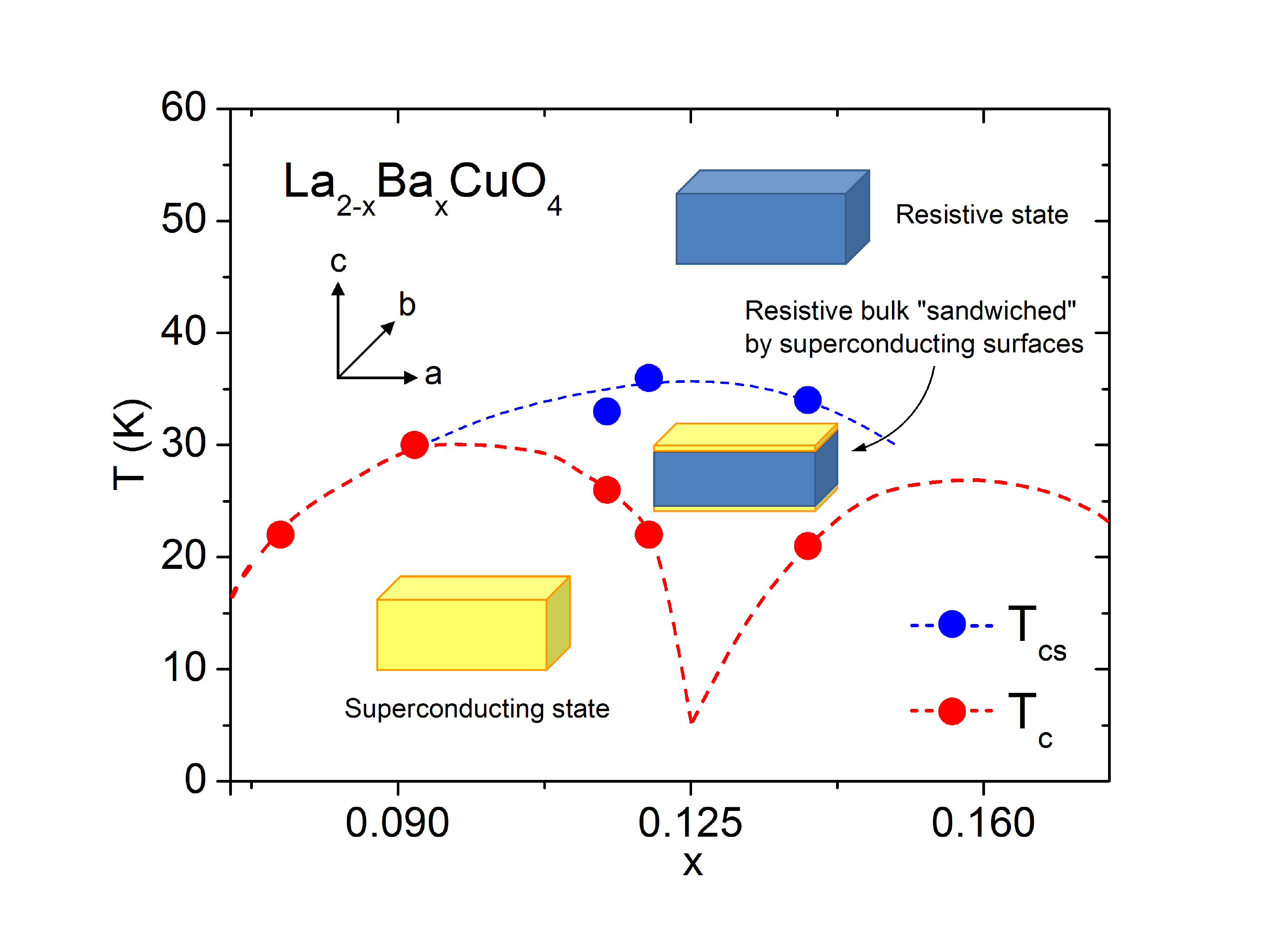}
\caption{$T_{cs}$ and $T_c$ as a function of $x$ from this work (solid circles). The dashed lines are drawn as a guide to the eye. Around $x=1/8$, the material exhibits the ``1/8-anomaly" where bulk superconductivity is suppressed \cite{Moodenbaugh88,Hucker11}{}. The higher $T_{cs}$ superconductivity was found in this 1/8-anomaly region. Also shown is a schematic illustrating the successive evolution of the sample as it is cooled to lower temperatures: first a fully resistive state for $T>T_{cs}$, then a resistive bulk ``sandwiched'' by superconducting surface regions for $T_c<T<T_{cs}$, and finally a fully superconducting state for $T<T_c$.} 
\end{figure*}

\bibliography{LBCO}
\end{document}

\title{Supplementary Information for ``Two Superconducting Transitions in Single Crystal La$_{2-x}$Ba$_x$CuO$_4$''}

\author{X. Y. Tee}
\affiliation{Division of Physics and Applied Physics, School of Physical and Mathematical Sciences, Nanyang Technological University, 637371 Singapore.}

\author{T. Ito}
\affiliation{National Institute of Advanced Industrial Science and Technology, Tsukuba, Ibaraki 305-8562, Japan.}

\author{T. Ushiyama}
\affiliation{National Institute of Advanced Industrial Science and Technology, Tsukuba, Ibaraki 305-8562, Japan.}

\author{Y. Tomioka}
\affiliation{National Institute of Advanced Industrial Science and Technology, Tsukuba, Ibaraki 305-8562, Japan.}

\author{I. Martin}
\affiliation{Materials Science Division, Argonne National Laboratory, Argonne, Illinois 60439, USA.}

\author{C. Panagopoulos}
\affiliation{Division of Physics and Applied Physics, School of Physical and Mathematical Sciences, Nanyang Technological University, 637371 Singapore.}
\affiliation{Department of Physics, University of Crete and FORTH, GR-71003 Heraklion, Greece}

\makeatletter
\newcommand*{\rom}[1]{\expandafter\@slowromancap\romannumeral #1@}
\makeatother
\let\oldhat\hat
\renewcommand{\hat}[1]{\oldhat{\mathbf{#1}}}

%\begin{document}
\maketitle

\section*{Resistivity and Thermopower Measurements}

Homemade apparatuses of high sensitivity (noise level $<\pm 5$ nV) were used for resistivity and thermopower measurements (Fig. S1-2). During the measurements, the sample was kept in high vacuum ($\sim 10^{-6}$ mbar) and inside a radiation shield which was thermally anchored to the sample stage. Silver wires of diameter 25 $\mu$m were used as lead wires to the sample. These minimized thermoelectric noise due to heat exchange between the sample and environment. A commercial Physical Property Measurement System (PPMS) was also used for resistivity measurement in an exchange gas environment. The results obtained in PPMS agreed with those using the homemade apparatus. 

Voltage and current contacts were made with silver paste DuPont 6838 baked in high purity O$_2$-flow at 450 $^{\circ}$C for 10 minutes. Contact resistance $<0.5$ $\Omega$ was typically achieved. In thermopower experiments, temperature gradient of typically $\sim 0.1$ K/mm was applied. Thermopower was measured at various temperature gradients to check for linearity.

\begin{figure}[H]
\begin{center}
\includegraphics[scale=0.1,trim = 30 0 0 0]{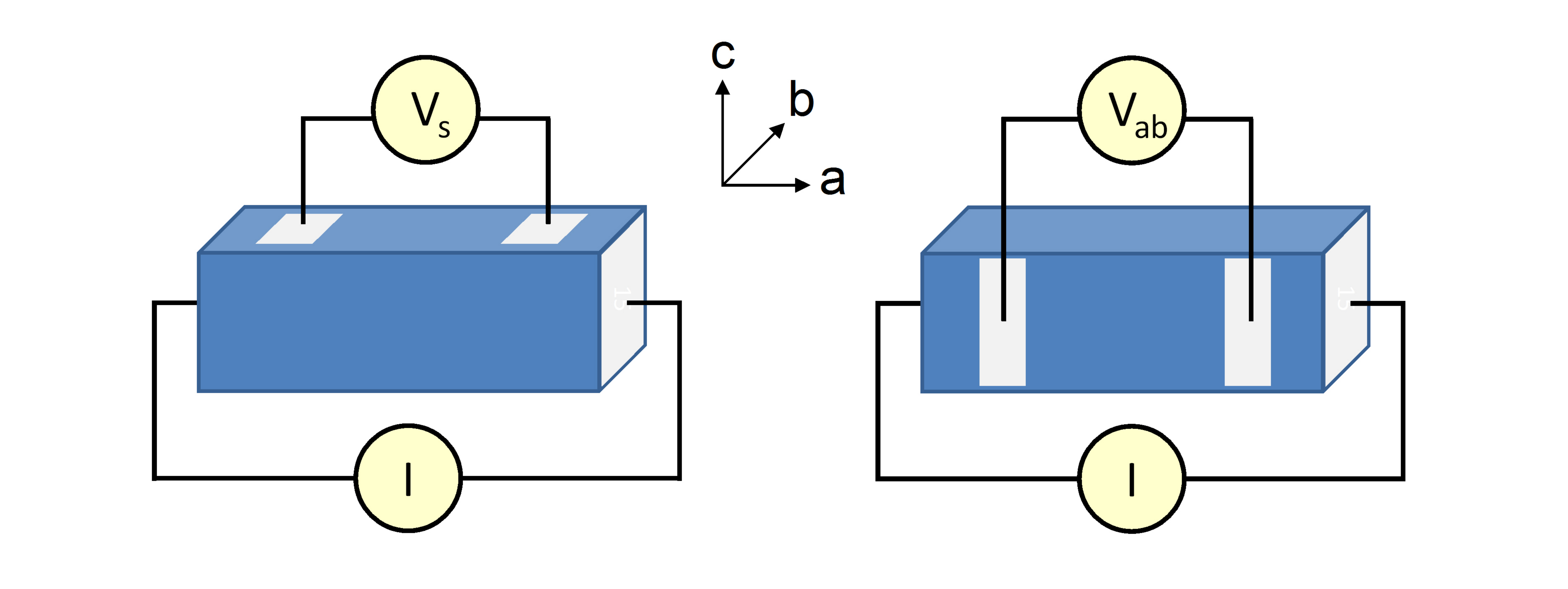}
\caption{Figure S1 $|$ Schematic contact configurations for resistivity measurement. Standard four-terminal method was used for resistivity measurement. Current $I$ was applied through cross-sectional area $A$ of the sample. Voltage contacts spaced at a distance $L$ were placed on the top of the sample (left panel) to measure the voltage drop $V_s$ on the surface, or on the side of the sample (right panel) to measure the voltage drop $V_{ab}$ in the bulk. The surface and bulk resistivities were calculated as $\rho_s=-V_s/I \times A/L$ and $\rho_{ab}=-V_{ab}/I \times A/L$, respectively.}
\end{center}
\end{figure}

\begin{figure}[H]
\begin{center}
\includegraphics[scale=0.13,trim = 30 0 0 0]{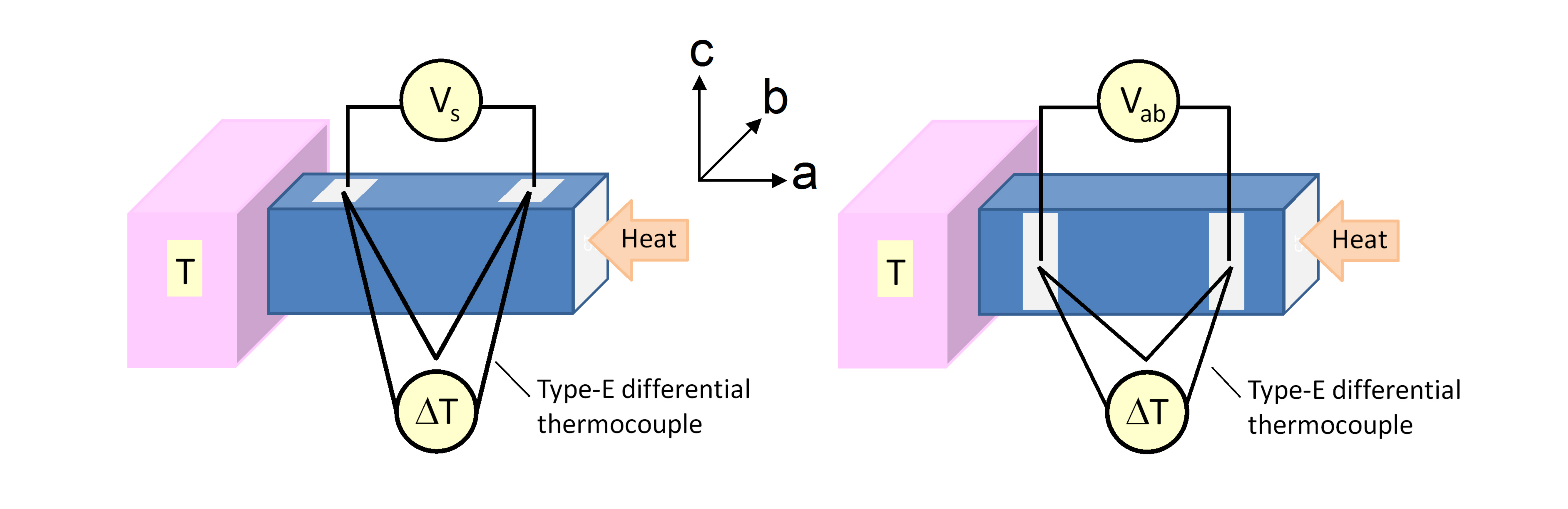}
\caption{Figure S2 $|$ Schematic contact configurations for thermopower measurement. One end of the sample is thermally anchored to the sample stage at temperature $T$. Heating power of typically $\sim 1$ mW was generated by a 1k$\Omega$ film heater and applied to the other end of the sample. Voltage contacts spaced at a distance $L$ were placed on the top of the sample (left panel) to measure the voltage drop $V_s$ on the surface, or on the side of the sample (right panel) to measure the voltage drop $V_{ab}$ in the bulk. Temperature difference $\Delta T$ between the voltage contacts was measured by a pair of type-E differential thermocouple wires of diameter 25 $\mu$m. The surface and bulk thermopowers were calculated as $S_s=-V_s/\Delta T$ and $S_{ab}=-V_{ab}/\Delta T$, respectively.}
\end{center}
\end{figure}

\section*{Sample Variation of $T_{cs}$}
The observed $T_{cs}$ is reproducible from sample to sample. Fig. S3 shows the case for LBCO-0.12 where $T_{cs}=36$ K is clearly seen in the two samples being measured.

\begin{figure}[H]
\begin{center}
\includegraphics[scale=0.35,trim = 30 0 0 0]{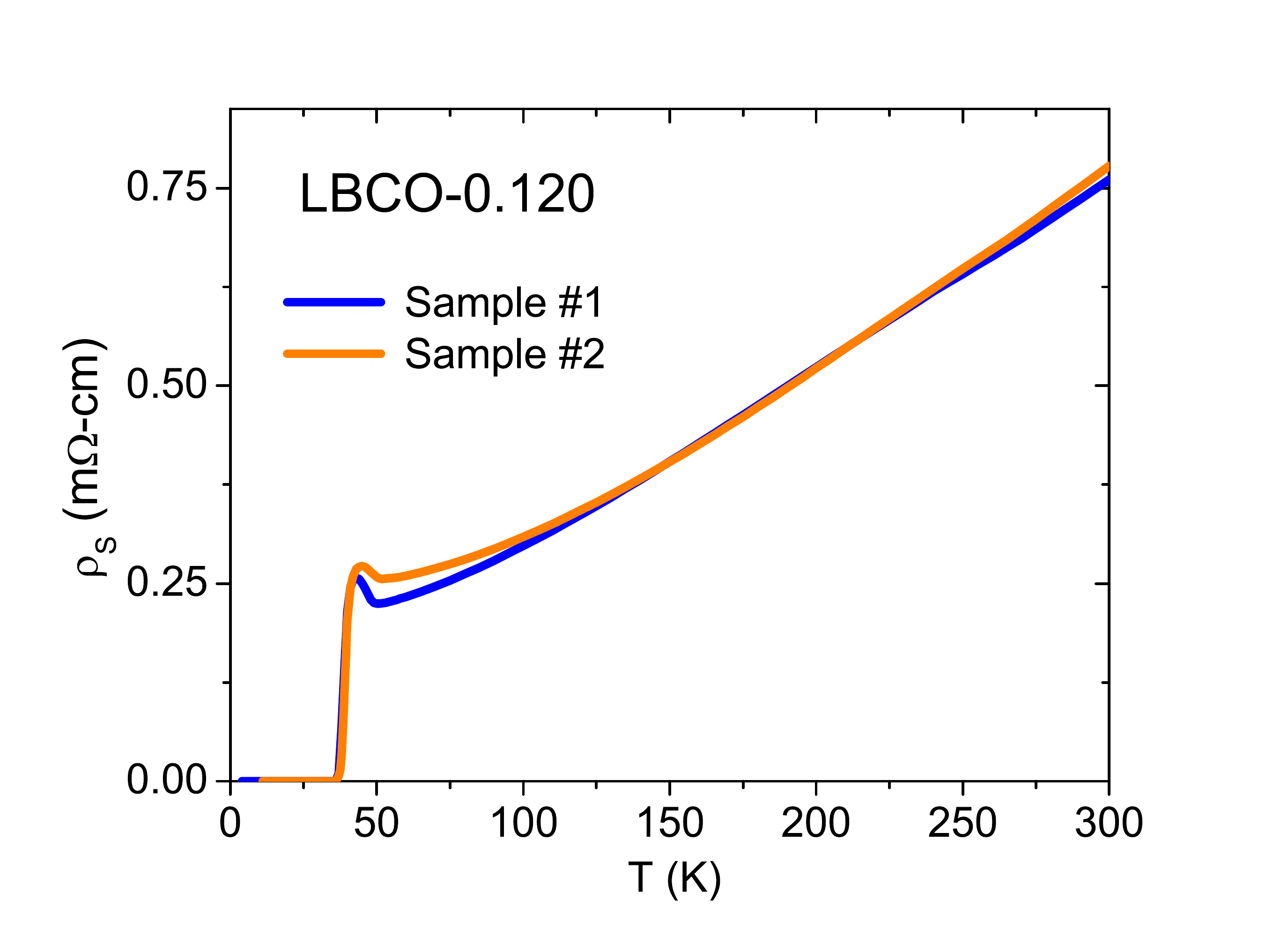}
\caption{Figure S3 $|$ Temperature dependence of surface resistivity in two samples of LBCO (x=0.12). Both samples show $T_{cs}$ = 36 K.}
\end{center}
\end{figure}

\newpage
\section*{Nernst Effect Measurements}
The Nernst effect is the generation of a transverse electric field $E||\hat{y}$ when the sample is subjected to an applied magnetic field $H||\hat{z}$ and temperature gradient $\nabla T||\hat{x}$.
In the mixed state of type-\rom{2} superconductors, vortex-flow produces a pronounced peak in the Nernst coefficient \cite{Josephson65,Huebener69,Hagen90,Zeh90,Xu00,Wang06}. In addition, there exists an onset temperature $T_{\nu}$ for the high temperature Nernst effect in cuprates that is strongly correlated with carrier doping level \cite{Wang06,Cyr09}. 
In La$_{2-x}$Sr$_x$CuO$_4$, for instance, a doping dependence of $~$500 K/hole was observed near x=0.12 \cite{Wang06}. 
Here, we measure the surface and bulk Nernst coefficient, $\nu_s$ and $\nu_{ab}$, of LBCO-0.120 (Fig. S4).
The measurements were performed with the same homemade apparatus used for thermopower measurements in a 16-Tesla superconducting magnet.

Figure S5 shows the temperature dependence of $\nu_s$ and $\nu_{ab}$.
At high temperatures, the two quantities overlap with $T_{\nu} \approx 110$ K. 
A common value of $T_{\nu}$ in the surface and bulk indicates uniform chemical doping throughout the sample.
The two quantities diverge, however, at lower temperatures ($T<T_d$).
Specifically, $\nu_s$ and $\nu_{ab}$ show a pronounced peak near $T_{cs}=36$ K and $T_c=22$ K, respectively, due to vortex-flow which is optimal just above the transition temperatures.
With decreasing temperature, the vortices get pinned causing a strong suppression on the coefficients. 
The vortex-pinning is also evidenced, for instance, in the field dependence of Nernst signal $e_s=\nu_s \times H$, where a minimum $H$ is required to trigger vortex-flow and attain a finite signal (Fig. S6).

In brief, our results indicate that surface superconductivity was observed also in Nernst effect measurements, in excellent agreement with the resistivity and thermopower measurements reported in the main text.

\begin{figure}[H]
\begin{center}
\includegraphics[scale=0.18,trim = 30 0 0 0]{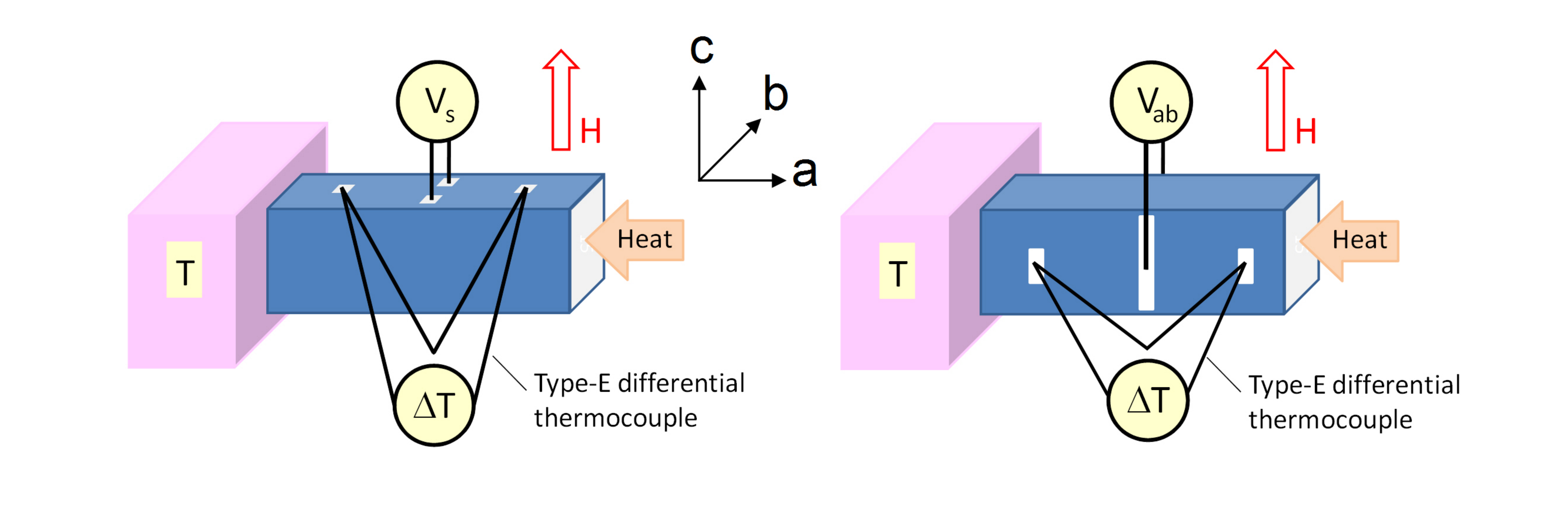}
\caption{Figure S4 $|$ Schematic contact configuration for the Nernst effect measurement. $H||\hat{z}$ was applied along the crystal c-axis. Heat was applied with the same heating configuration used for thermopower measurement. Voltage contacts spaced at a distance $W||\hat{y}$ were placed on the top of the sample (left panel) to measure the voltage drop $V_s$ on the surface, or on the side of the sample (right panel) to measure the voltage drop $V_{ab}$ in the bulk. Temperature difference $\Delta T$ was measured by a pair of type-E differential thermocouple wires spaced at $L||\hat{x}$ on the same side with the voltage contacts. 
The surface and bulk Nernst coefficients are calculated as $\nu_{s}=V_s/W \times L/\Delta T \times 1/H$ and $\nu_{ab}=V_{ab}/W \times L/\Delta T \times 1/H$, respectively.
The sign of $\nu_s$ and $\nu_{ab}$ is defined by the vortex-flow convention for the Nernst effect \cite{Josephson65,Wang06}{}. To remove the unavoidable pick-up of thermopower (field-symmetric) due to misalignment of voltage contacts, $H$ was applied in both directions so that only the field-antisymmetric part of the data is taken.}
\end{center}
\end{figure}

\begin{figure}[H]
\begin{center}
\includegraphics[scale=0.37,trim = 30 0 0 0]{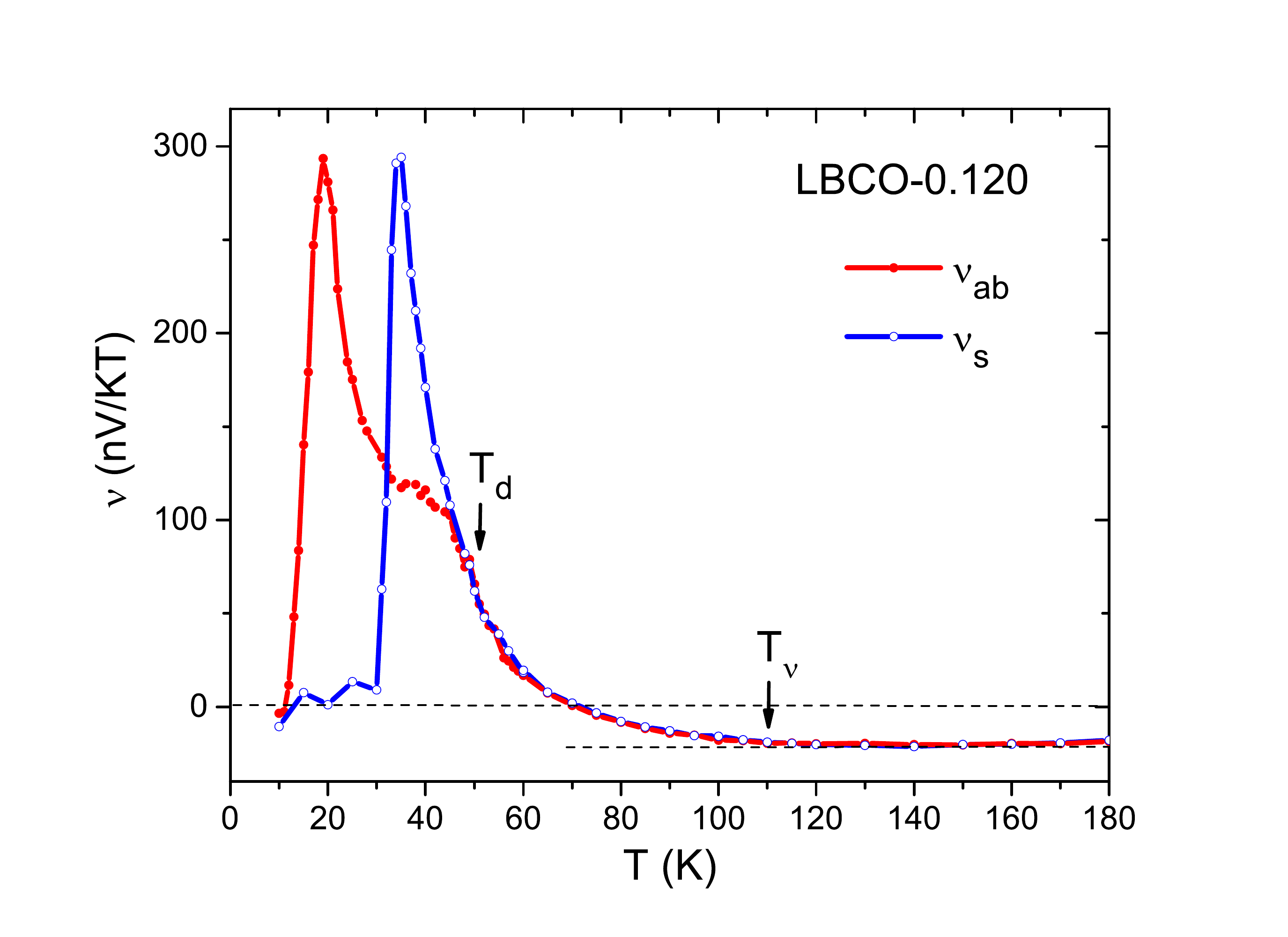}
\caption{Figure S5 $|$ Temperature dependence of \boldmath$\nu_s$ and \boldmath$\nu_{ab}$. The measurements were performed with $H=1$ T for $T<60$ K. At higher temperatures where $\nu_s$ and $\nu_{ab}$ are weak and field-independent, $H=14$ T was used to achieve a better signal-to-noise ratio. 
$\nu_s$ and $\nu_{ab}$ overlap in the normal state with the same $T_{\nu}$, but diverge below $T_d$ due to the onset of surface superconductivity. Pronounced peaks in the two quantities are due to optimal vortex-flow near the superconducting transitions.}
\end{center}
\end{figure}

\begin{figure}[H]
\begin{center}
\includegraphics[scale=0.37,trim = 30 0 0 0]{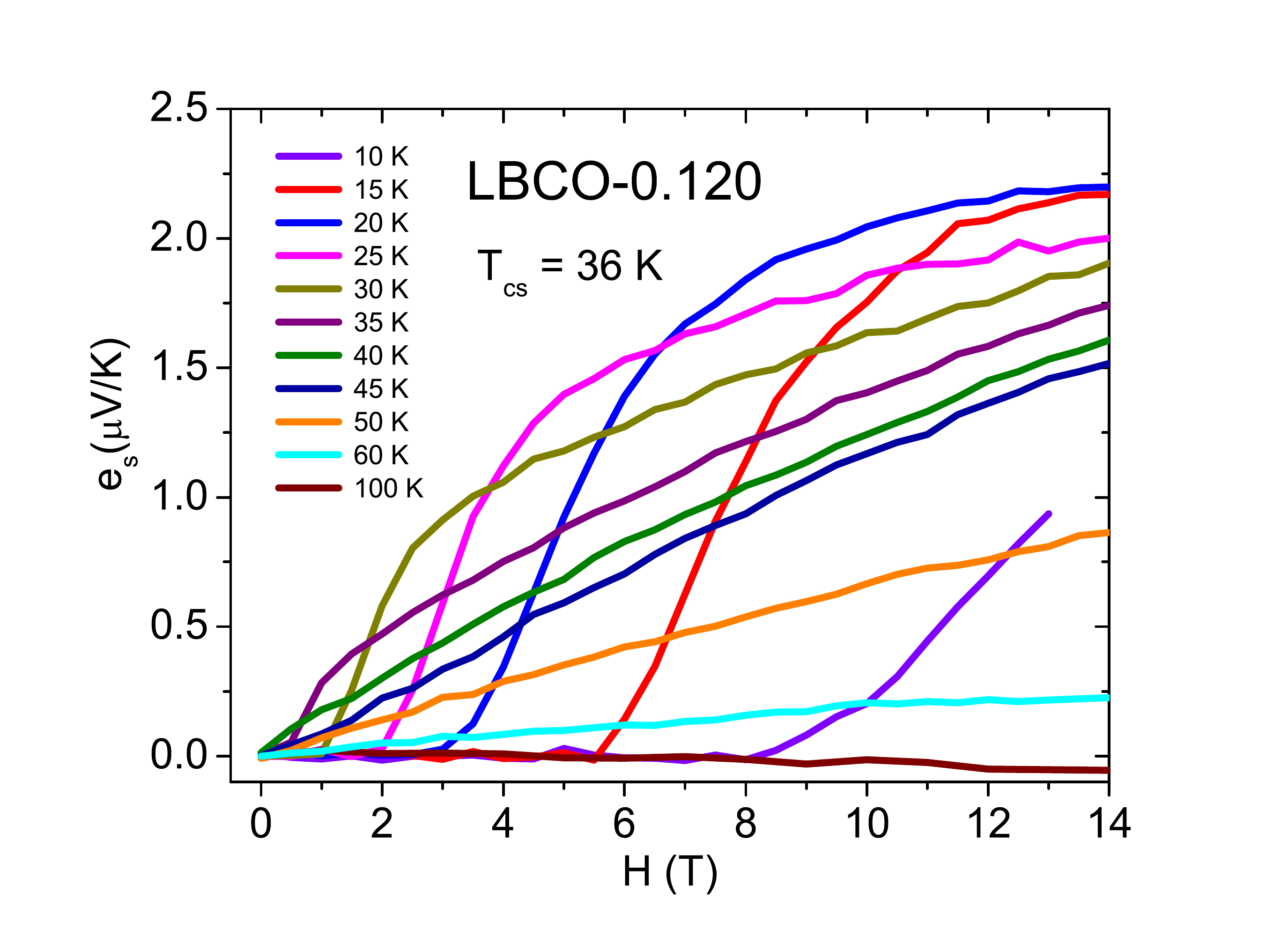}
\caption{Figure S6 $|$ Field dependence of \boldmath$e_s$ at various temperatures. At high temperatures, curves of $e_s(H,T)$ show the linear $H$-dependence generally observed in underdoped cuprates \cite{Xu00,Wang06}. Below $T_{cs}$, the curves show nonlinear $H$-dependence which is characteristic of vortex pinning where a minimum $H$ is required to re-activate the vortex-flow to attain a finite value of $e_s(H,T)$.}
\end{center}
\end{figure}

\newpage
\section*{Resistivity Scaling}
Here, we will demonstrate that a scaling does exist for $\rho_s$  and $\rho_{ab}$ at temperatures higher than $T_{cs}$. For a homogeneous sample, the normal-state resistivity should be uniform throughout the sample. Therefore, the surface resistivity should satisfy 
\[\rho_s^n=A\rho_{ab}\]
Where the superscript denotes normal-state property, $A\approx 1$ is a scaling factor that does not change with temperature.

We further note that in reality, the surface may not be perfectly parallel to the ab-plane. This could happen, for instance, due to a small but unavoidable misalignment in the cutting of samples. Subsequent surface polishing may also modify the alignment. Therefore, the sample c-axis makes an angle $\theta_c$ with respect to the surface normal. Due to the strong anisotropy in resistivity, e.g. $\rho_c/\rho_{ab}\approx10^{-3}$ in the case of LBCO-0.120 near $T=300$ K (Fig. S7d), a small $\theta_c$ will cause a sizable pickup of $\rho_c$ when measuring $\rho_s$. Therefore, in the normal state, the observed values of surface resistivity, $\rho_s^{obs}$, may be described as

\begin{align*}
\rho_s^{n,obs} &= \rho_s^n+B\rho_c \\
&= A\rho_{ab}+B\rho_c 
\end{align*}

where $B$ is the pickup coefficient which is independent of temperature.

Figure S7a-e show that for all our samples, $\rho_s$ can be fitted by $\rho_s^{obs}$ over a wide range of temperatures. At low temperatures, the onset of surface superconducting fluctuations makes the surface more conductive than the normal-state counterpart, i.e. $\rho_s^n<A\rho_{ab}$. Furthermore, proximity effect \cite{Bozovic94,Bozovic04} shorts c-axis resistivity within the surface layer which consists of several copper-oxide planes. As a result, $\rho_s$ deviates downward from $\rho_s^{obs}$  and eventually vanishes at $T_{cs}$.

We note that for $x=0.115$ (Fig. S7c), in run $1$, a large pickup coefficient $B=14.5\times 10^{-4}$ causes a strong deviation of $\rho_s$ from $\rho_{ab}$. In run $2$, however, measurement on the re-polished surface contains a smaller $\rho_c$ pickup with $B=3\times 10^{-4}$. The important point is that while polishing modifies the surface alignment and hence the degree of $\rho_c$ pickup, it does not affect $T_{cs}$. For $x=0.139$ (Fig. S7e), where surface $T_{cs}$ is also observed, the $\rho_c$ pickup is in fact zero. Specifically, excellent scaling  $\rho_s=1.6\rho_{ab}$ is observed for $T>40$ K. These observations indicate the pickup is a side effect due to imperfect experimental condition, which can in principle be completely removed, and not relevant to the enhancement on $T_{cs}$.

Thus the seeming absence of scaling between $\rho_s$ and $\rho_{ab}$ is most likely due to a small but unavoidable pickup of $\rho_{c}$ in the former.

\begin{figure}[H]
\begin{center}
\includegraphics[scale=0.13,trim = 30 0 0 0]{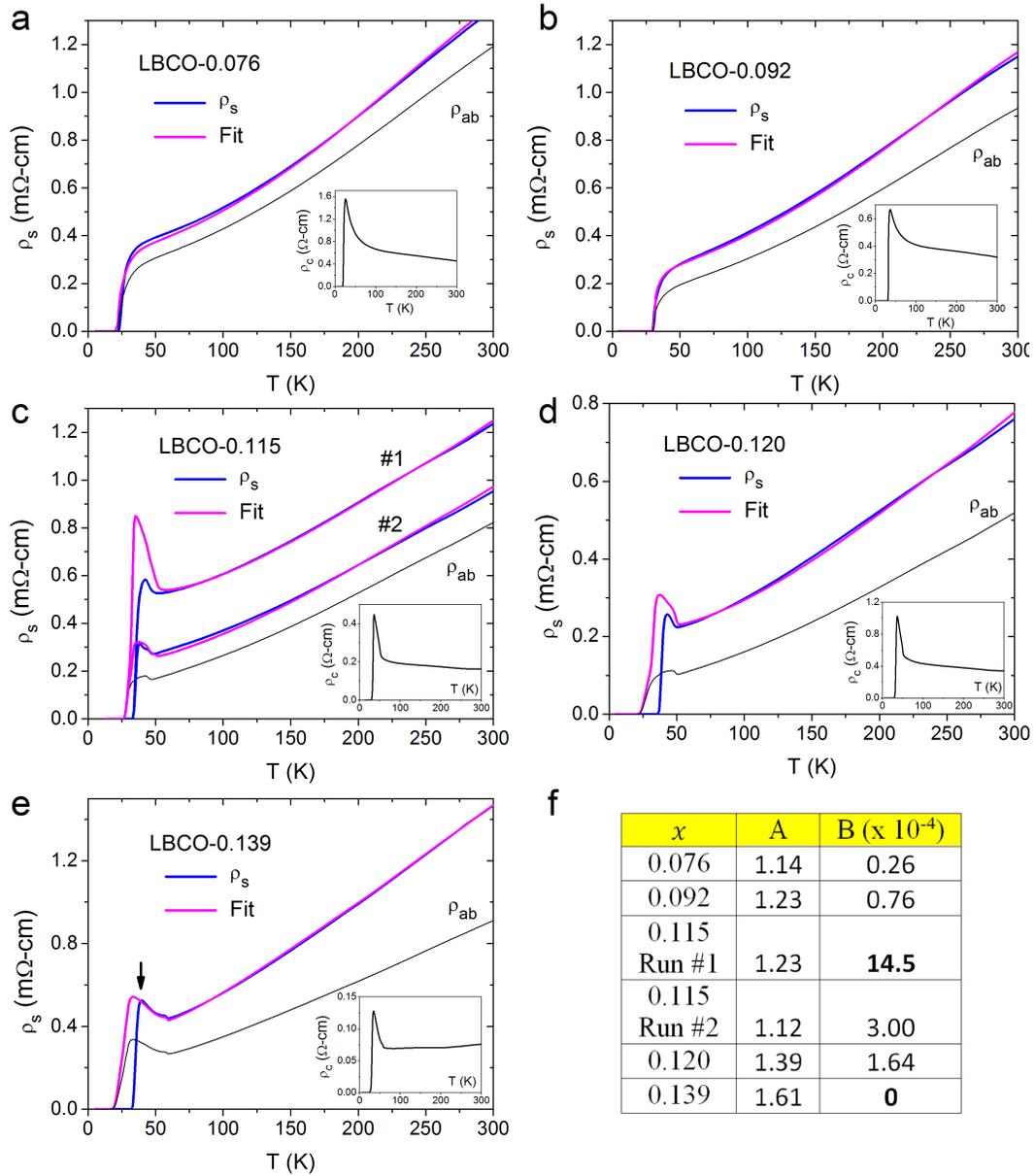}
\caption{Figure S7 $|$ Scaling of \boldmath$\rho_s$ and \boldmath$\rho_{ab}$ at temperatures higher than $T_{cs}$. (a)-(e) $\rho_s$ for various doping levels $x$ are fit by $\rho_s^{n,obs}=A\rho_{ab}+B\rho_c$. Also shown are the temperature dependence of $\rho_{ab}$ and $\rho_c$ (inset). (f) Fitting coefficients $A$ and $B$ used in (a)-(e).}
\end{center}
\end{figure}

\newpage
\section*{Resistance Tomography}
The basic idea is to measure the resistance of different regions of the sample and check for consistency. To do this, multiple contacts were placed with silver paste on the sides and top of the sample (four contacts on each) for bulk and surface measurements, respectively (Fig. S8). We measured the resistance across pairs of contacts $i$ and $j$, defined as $R_{ij}=V_{ij}/I_{ij}$ and $R_{s(ij)}=V_{ij}/I_{ij}$ for the bulk and the surface configurations, respectively, by applying current $I_{ij}$ and sensing the voltage drop $V_{ij}$ strictly through the same pair of contacts.
In this setting, $R_{ij}$ and $R_{s(ij)}$ contain two terms, namely the resistance of the local region sensed by the current path, and the contact resistances which as we demonstrate below, can be removed.
\\

Figures S7a-b show the temperature dependence of $R_{ij}$ and $R_{s(ij)}$ for all combinations of surface and bulk contacts of $\{i,j\}$. $R_{ij}$ and $R_{s(ij)}$ show a rapid drop at the respective $T_c$ and $T_{cs}$ due to the \textit{two distinct} superconducting transitions. At lower temperatures, both $R_{ij}$ and $R_{s(ij)}$ are dominated by the contact resistances which show semiconducting-like temperature dependence. The results indicate that the bulk and surface superconductivity are consistently observed in different local regions of the bulk and surface, respectively. It is also important to note that $T_{cs}$ is observed exclusively in the surface measurements, thus confirming the confinement of surface superconductivity to a thin surface layer.
\\

Next, we remove the contact resistances for a more rigorous test of sample homogeneity. We calculate the combination $R_{ij,kl} = R_{ij} + R_{kl}$ and $R_{s(ij,kl)} = R_{s(ij)} + R_{s(kl)}$ (Figs. S9c-d). Note that the sum of all the contact resistances is reproducible over measurement runs, as seen in both sets of $R_{ij,kl}$ and $R_{s(ij,kl)}$ overlapping below the superconducting transition temperatures. This fact enabled us to cancel out the contact resistance, by calculating $\Delta R_{ij,kl}$ and $\Delta R_{s(ij,kl)}$ which are the differences between two combinations (e.g. $R_{14,23}-R_{12,34}$). Figures S9e-f show the main results where $\Delta R_{ij,kl}$ and $\Delta R_{s(ij,kl)}$ are rescaled to the bulk and surface resistances measured by the standard four-terminal method, $R_{ab}$ and $R_s$, respectively. The excellent scaling observed here, indicates consistency in the resistance measured in different parts of the sample. This confirms the sample homogeneity within the bulk and the surface.

\begin{figure}[H]
\begin{center}
\includegraphics[scale=0.1,trim = 30 0 0 0]{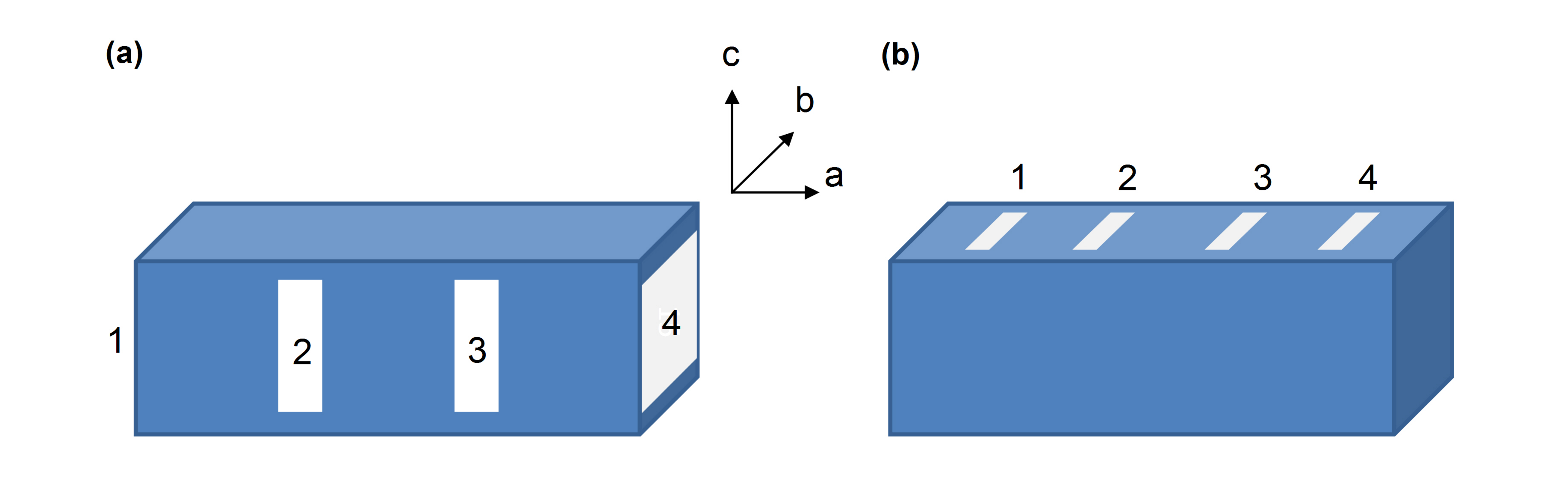}
\caption{Figure S8 $|$ Schematic contact configurations for ``resistance tomography" experiments on LBCO-0.120. Four contacts numbered $i=$1, 2, 3, and 4 were placed on the sides or top of the sample for (a) the bulk and (b) the surface measurements, respectively. In each configuration, current $I_{ij}$ was applied through two contacts $i$ and $j$ and the corresponding voltage drop $V_{ij}$ was measured, for all six combinations of $\{i,j\}$. The regional resistance across any two contacts is defined as $R_{ij}=V_{ij}/I_{ij}$ for the bulk measurements, and $R_{s(ij)}=V_{ij}/I_{ij}$ for the surface measurements.}
\end{center}
\end{figure}

\begin{figure}[H]
\begin{center}
\includegraphics[scale=1.05,trim = 0 0 0 0]{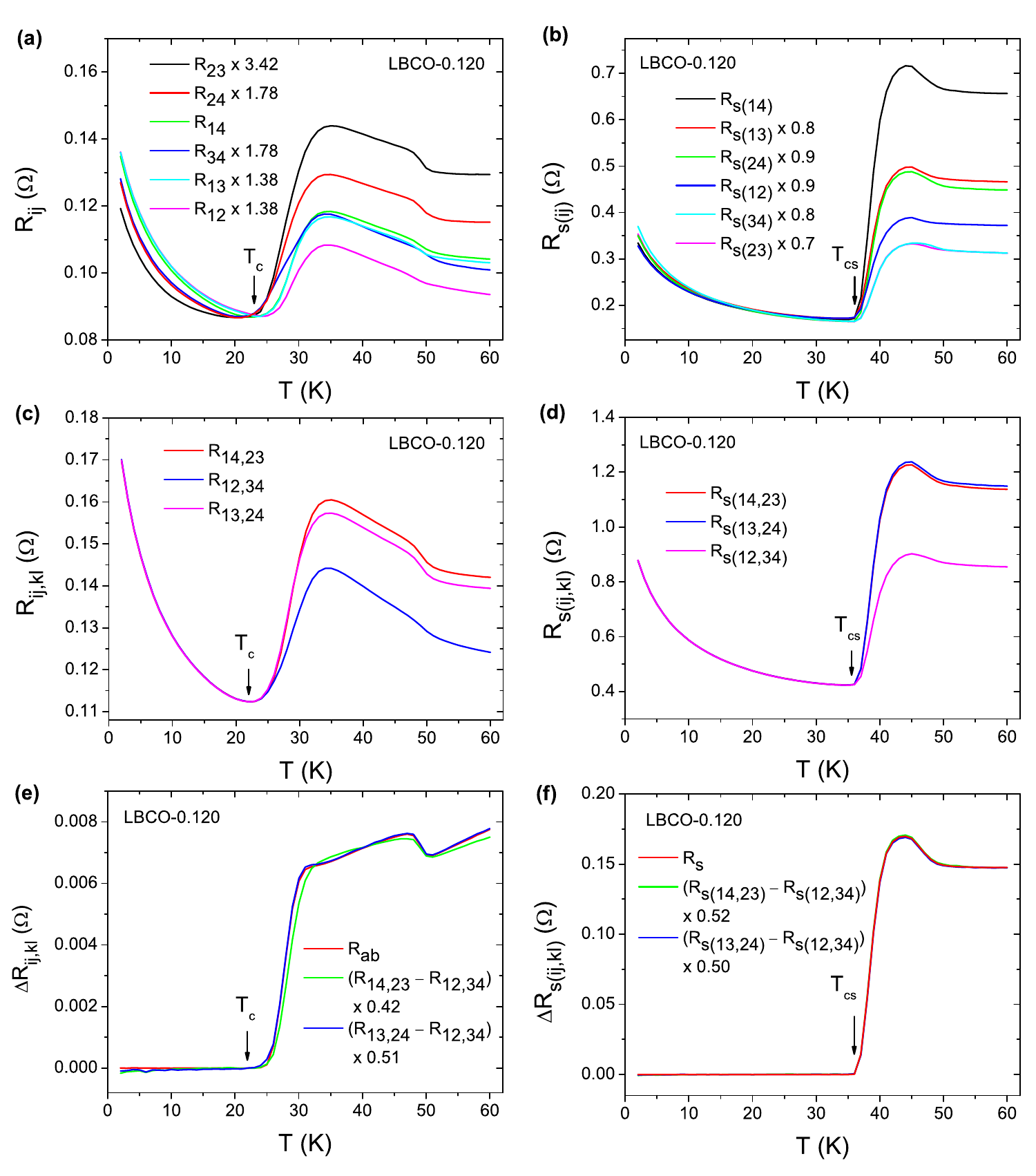}
\caption{Figure S9 $|$ Resistance tomography of LBCO-0.120. a-b, Bulk $R_{ij}$ and surface $R_{s(ij)}$ as defined in the supplementary text, are rescaled to demonstrate the general trends. Both sets of $R_{ij}$ and $R_{s(ij)}$ show rapid drops at the respective $T_{c}$ and $T_{cs}$, consistent with the \textit{two distinct} superconducting transitions. At lower temperatures, semiconducting-like contact resistances dominate. The applied current was 1 mA in each measurement run. c-d, $R_{ij,kl}=R_{ij}+R_{kl}$ and $R_{s(ij,kl)}=R_{s(ij)}+R_{s(kl)}$ are calculated. The overlapped part below the respective $T_{c}$ and $T_{cs}$ is the sum of all the contact resistances for each configuration. e-f, The differential values of $\Delta R_{ij,kl}$ and $\Delta R_{s(ij,kl)}$ are calculated in order to cancel out the contact resistances. $R_{14,23}-R_{13,24}$ and $R_{s(14,23)}-R_{s(13,24)}$ are omitted due to their relatively small values. Clearly, $\Delta R_{ij,kl}$ and  $\Delta R_{s(ij,kl)}$ can be rescaled to $R_{ab}$ and $R_s$, respectively, which are the bulk and surface resistances measured by the standard four-terminal method.}
\end{center}
\end{figure}

\section*{Berezinskii-Kosterlitz-Thouless transition}

Two-dimensional (2D) superconductivity can be described by Berezinskii-Kosterlitz-Thouless (BKT) transition  \cite{Bere72,Koster73,Beasley79,Halperin79}.
Near the transition, the I-V curves are expected to obey $V\propto I^{\alpha}$ with $\alpha=3$ at $T_{BKT}$, and the resistivity follows the temperature dependence $\rho \propto \exp(-b/t)$, where $b$ is material parameter, and  $t=T/T_{BKT}-1$. 
Here, we analyse the surface I-V curves and  $\rho_s$ of LBCO-0.120, both measured with the same contact configuration (Fig. S1).
Figure S10 presents our results fitted by BKT with $T_{BKT}=36.6$ K and $b=0.996$.
The fits reveal 2D-like feature of the surface superconductivity.
We note that similar BKT behaviours were also observed in the \textit{bulk} of LBCO-1/8 \cite{Li07}. 
However, the bulk parameters $T_{BKT}=16.3$ K and $b=2.7$ in their case are considerably different from our values, pointing to a distinction between surface and bulk.

In the measurements with surface contacts, we do not anticipate significant influence of the bulk when the surface becomes superconducting. This is because the surface conductance dominates the bulk counterpart. Thus we can expect the superconducting behavior of the surface, notably the BKT scaling, to be unaffected by the proximity to underlying metal. The significance of the BKT analysis is that it indicates surface superconductivity is confined to a thickness not exceeding the superconducting coherence length $\xi$ which is comparable to the atomic lattice constant of this material. However, it could emerge from a set of superconducting planes with no mutual phase coherence but an overall thickness larger than $\xi$ \cite{Li07}. With the present experiment we cannot distinguish between the two scenarios.

\begin{figure}[H]
\begin{center}
\includegraphics[scale=0.7,trim = 30 0 0 0]{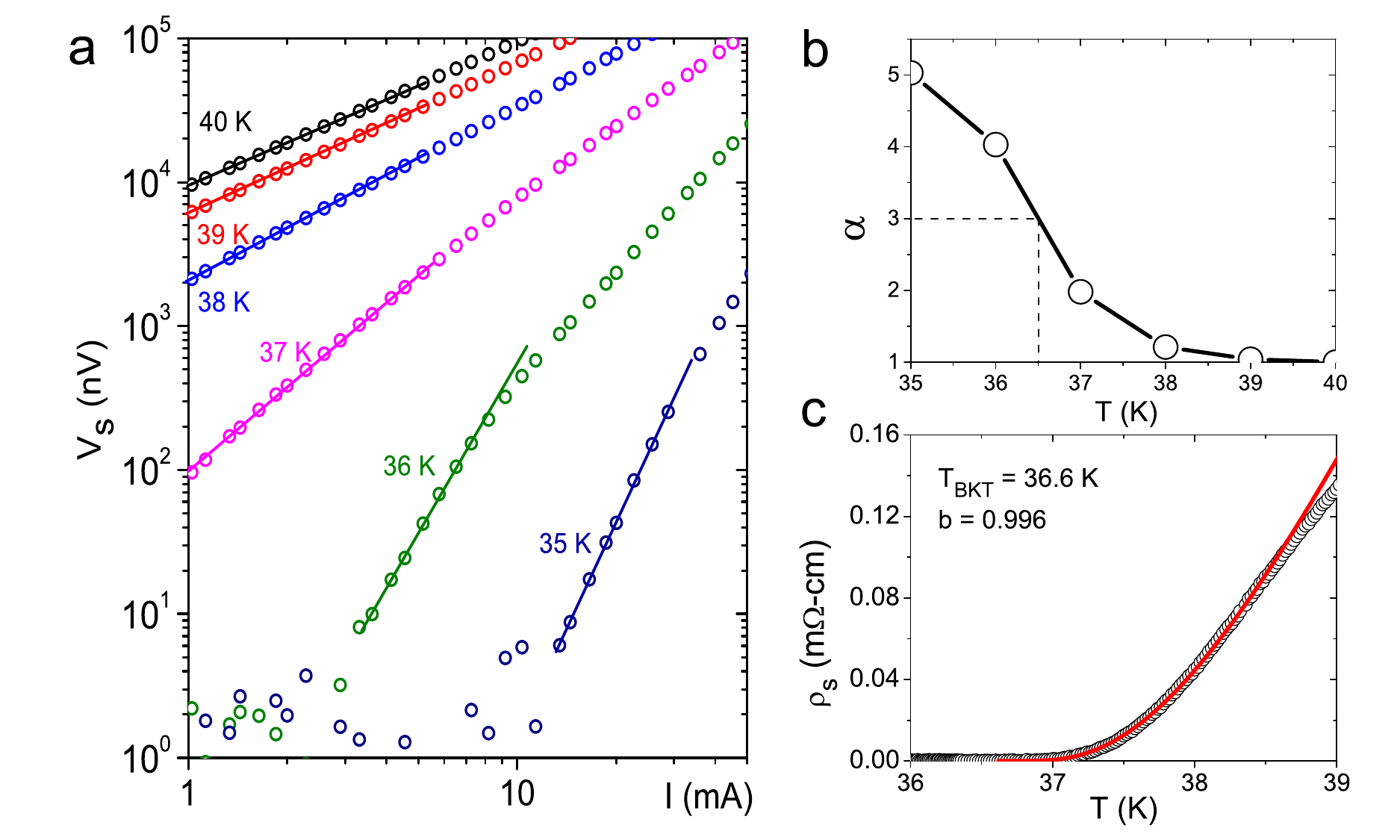}
\caption{Figure S10 $|$ Surface I-V characteristics and $\rho_s$ of LBCO-0.120 fitted by BKT. a, Surface I-V curves on a logarithmic scale, measured with the $\rho_s$ contact configuration at different temperatures near $T_{cs}$. The solid lines are fits to the power law $V\propto I^{\alpha}$. b, The exponent $\alpha$ as a function of temperature, reaching a value of 3 at $T_{BKT}\approx 36.5$ K. c, Near the superconducting transition, $\rho_s$ is well fitted by $\rho \propto \exp(-b/t)$ (solid line) with $b=0.996$ and $T_{BKT}=36.6$ K.}
\end{center}
\end{figure}

\newpage
\section*{Superconducting Surface Thickness}

Here, we analyze the critical current measurement for an estimate of the superconducting surface thickness. Fig. S10 shows the applied current required to destroy the surface superconductivity in LBCO-0.120 in the surface contact configuration (inset of Fig. S10). At zero temperature, the extrapolated value is $I_c(0)\approx 2500$ mA.
However, the actual current flowing through the surface is $I^s_c=sI_c$ where $s$ is the shunting factor due to sample geometry and resistivity anisotropy, which is estimated to be 2.5 in our case (for detailed calculation, see supplementary section ``Resistor Network Model"). Thus the zero-temperature value of the \textit{surface} critical current is:
\begin{align*}
I^s_c(0)&\approx 0.025 \times 2500 \text{ mA} \\
&\approx 62.5 \text{ mA}
\end{align*}
\begin{figure}[H]
\begin{center}
\includegraphics[scale=0.1,trim = -60 0 0 0]{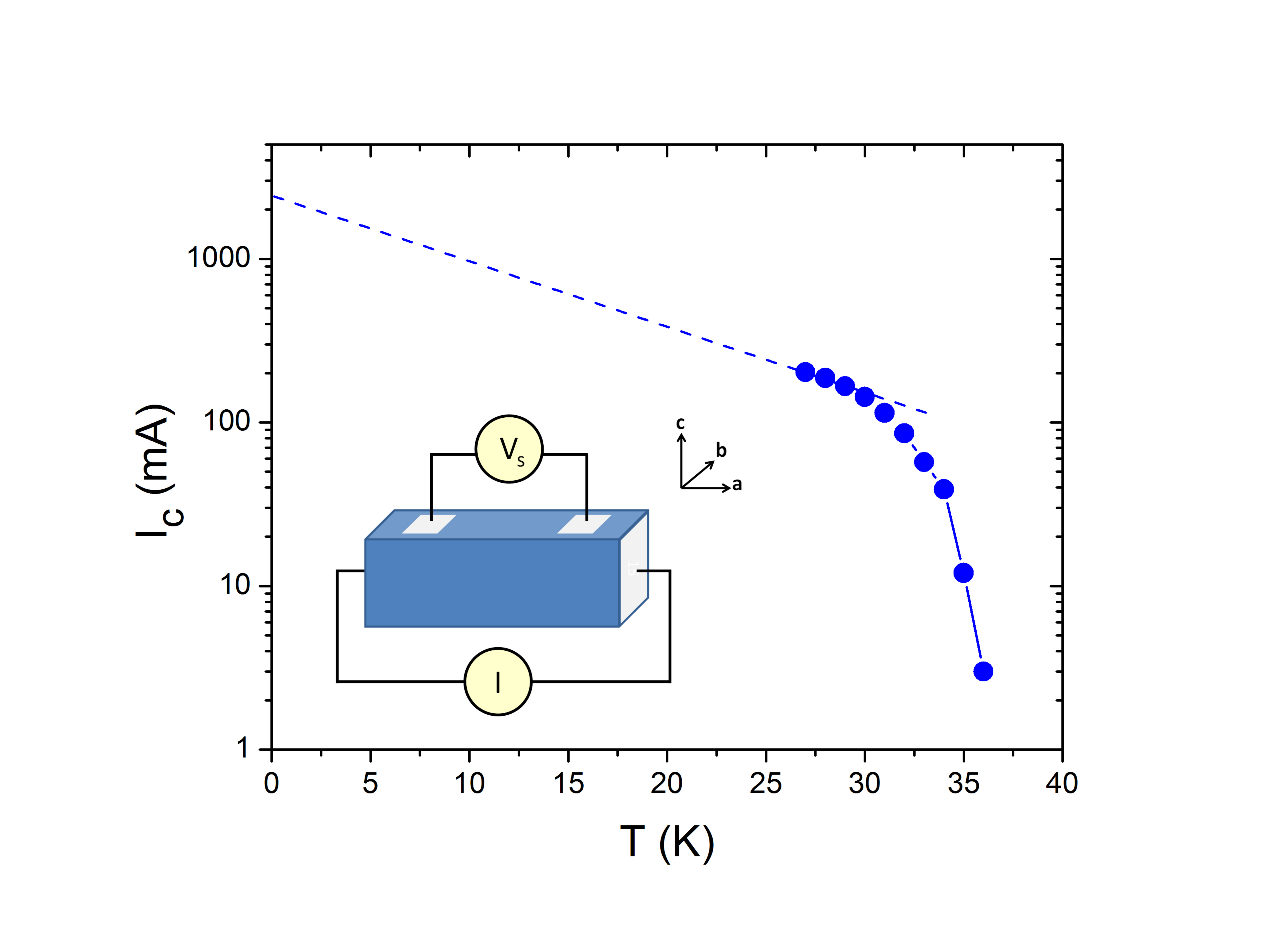}
\caption{Figure S11 $|$ Temperature dependence of the applied current $I_c$ required to destroy the surface superconductivity in LBCO-0.12 ($T_{cs}=36$ K). The zero-temperature value is estimated to be 2500 mA by extrapolation. The inset shows the surface contact configuration used in this measurement.}
\end{center}
\end{figure} 

The surface critical current density is calculated as $J^s_c(0)=I^s_c(0)/(w\times d)$ where $w$ is the width of sample, and $d$ the thickness of the superconducting surface. Given $I^s_c(0)\approx 62.5$ mA and $w=0.5$ mm in our case, $J^s_c(0)$ for a range of $d$ are estimated and displayed in Table T1:

\begin{figure}[H]
\begin{center}
\includegraphics[scale=0.3,trim = -60 0 0 0]{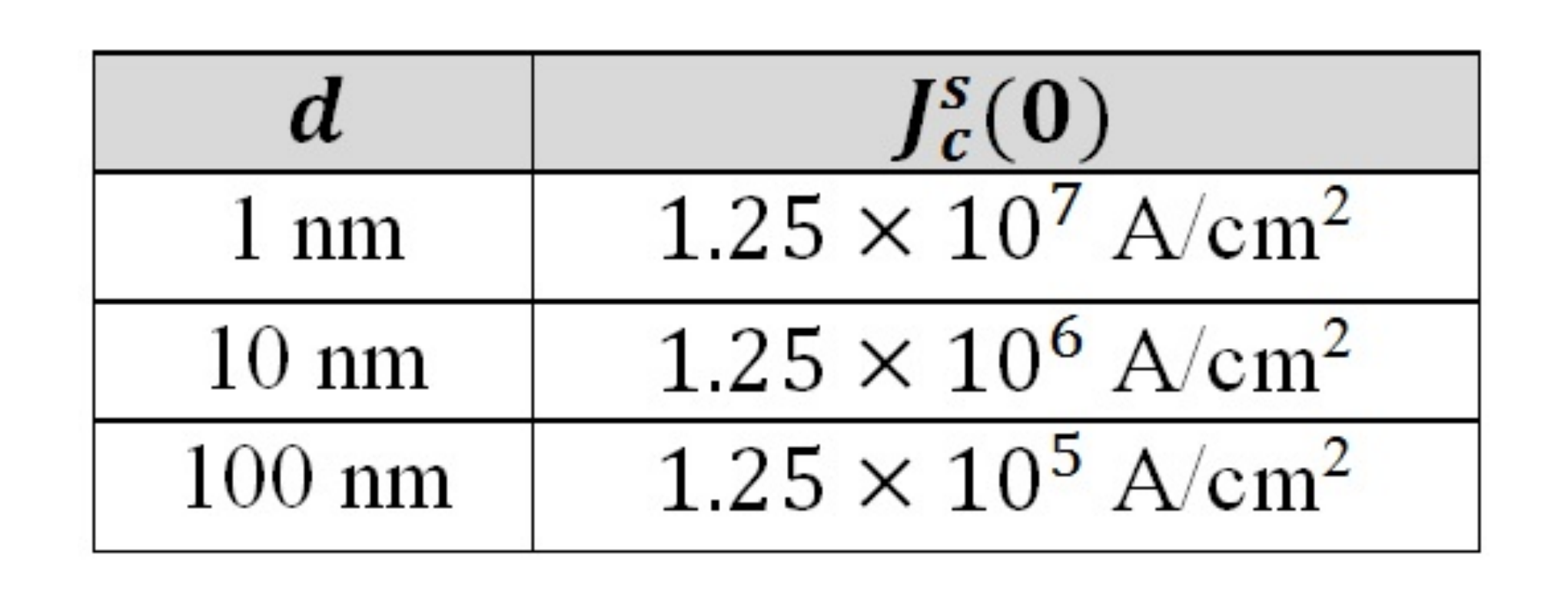}
\caption{Table T1 $|$ Surface critical current density $J^s_c(0)$ estimated for various possible values of the superconducting surface thickness $d$.}
\end{center}
\end{figure} 

These values are compared to the optimally doped La$_{2-x}$Sr$_x$CuO$_4$ which is the sister compound of La$_{2-x}$Ba$_x$CuO$_4$ and importantly has a $T_c\approx 36$ K similar to the surface $T_{cs}$ of our sample. The cases of $d=1$ nm with $J^s_c(0)\approx 1.25\times 10^7$ A/cm$^2$ and $d=10$ nm with $J^s_c(0)\approx 1.25\times 10^6$ A/cm$^2$ are closest to the reported value of $J_c(0)\approx 6\times 10^7$ A/cm$^2$ in optimally doped La$_{2-x}$Sr$_x$CuO$_4$ \cite{Wen03}. Hence, even though from the present measurements we cannot directly extract the thickness of the superconducting layer, the result that we obtain is not inconsistent with superconductivity confined to a thin surface layer on the nm scale.

\newpage
\section*{Resistor Network Model}
Here we explicitly consider a resistor network model to analyse resistivity measurements in the presence of a superconducting surface layer. As we shall show, surface superconductivity is ineffective in shunting the current from the bulk.

Since the sample can be viewed as repetition of the ac-plane along the b-axis, it suffices to consider the current flow in the ac-plane which essentially captures the shunting effect of the surface. The ac-plane can be described as an anisotropic resistive plane with dimensions $L_x$ and $L_y$ along the x and y directions, respectively, where the x and y directions are set parallel to the a- and c-axes of the sample. In the absence of magnetic field, the current density $\hat{J}$ and electric field $\hat{E}$ are related by a diagonal conductivity tensor $\sigma=\left[\begin{array}{cc} \sigma_{xx} & 0\\0 & \sigma_{yy} \end{array}\right]$ such that

\[
\hat{J}=\sigma\hat{E}=\textrm{\ensuremath{\hat{x}}}\sigma_{xx}E_{x}+\textrm{\ensuremath{\hat{y}}}\sigma_{yy}E_{y}
\]
Substitute $\sigma_{xx}=\sigma_{ab}$, $\sigma_{yy}=\sigma_{c}$, and $\hat{E}=-\nabla\phi$ into the equation, we get

\[
\hat{J}=-\textrm{\ensuremath{\hat{x}}}\sigma_{ab}\partial_{x}\phi+\textrm{\ensuremath{\hat{y}}}\sigma_{c}\partial_{y}\phi
\]
Since there is no charge accumulation, $\nabla\cdot\hat{J}=0$,

\[
\Rightarrow\sigma_{ab}\partial_{x}^{2}\phi+\sigma_{c}\partial_{y}^{2}\phi=0
\]
Defining $\tilde{y}=y\sqrt{\sigma_{ab}/\sigma_{c}}$, we get

\[
\partial_{x}^{2}\phi+\partial_{\tilde{y}}^{2}\phi=0
\]

Thus we have mapped the system into an isotropic resistive plane with
the new dimensions $L_{x}\times L_{\tilde{y}}$, where $L_{\tilde{y}}=\sqrt{\sigma_{ab}/\sigma_{c}}L_{y}$.

To perform numerical analysis, we discretise the isotropic resistive plane into a rectangular mesh of $N_{x}\times N_{y}$ sites, where $N_{x}\propto L_{x}$ and $N_{y}\propto L_{\tilde{y}}$ (Fig. S12).
The sites are interconnected through resistors $R$, except for those at the top and bottom boundaries (i.e. the surfaces), which are connected to the inner sites through $R$ but are horizontally interconnected
through resistors $R_{s}$. $R_{s}$ represents the surface layer
and is set to zero at temperatures below $T_{cs}$. In addition, we
model the current contacts as contact resistors $R_{c}$ connected
to the sites on the left and right boundaries.

\begin{figure}[H]
\begin{center}
\includegraphics[scale=0.2,trim = -60 0 0 0]{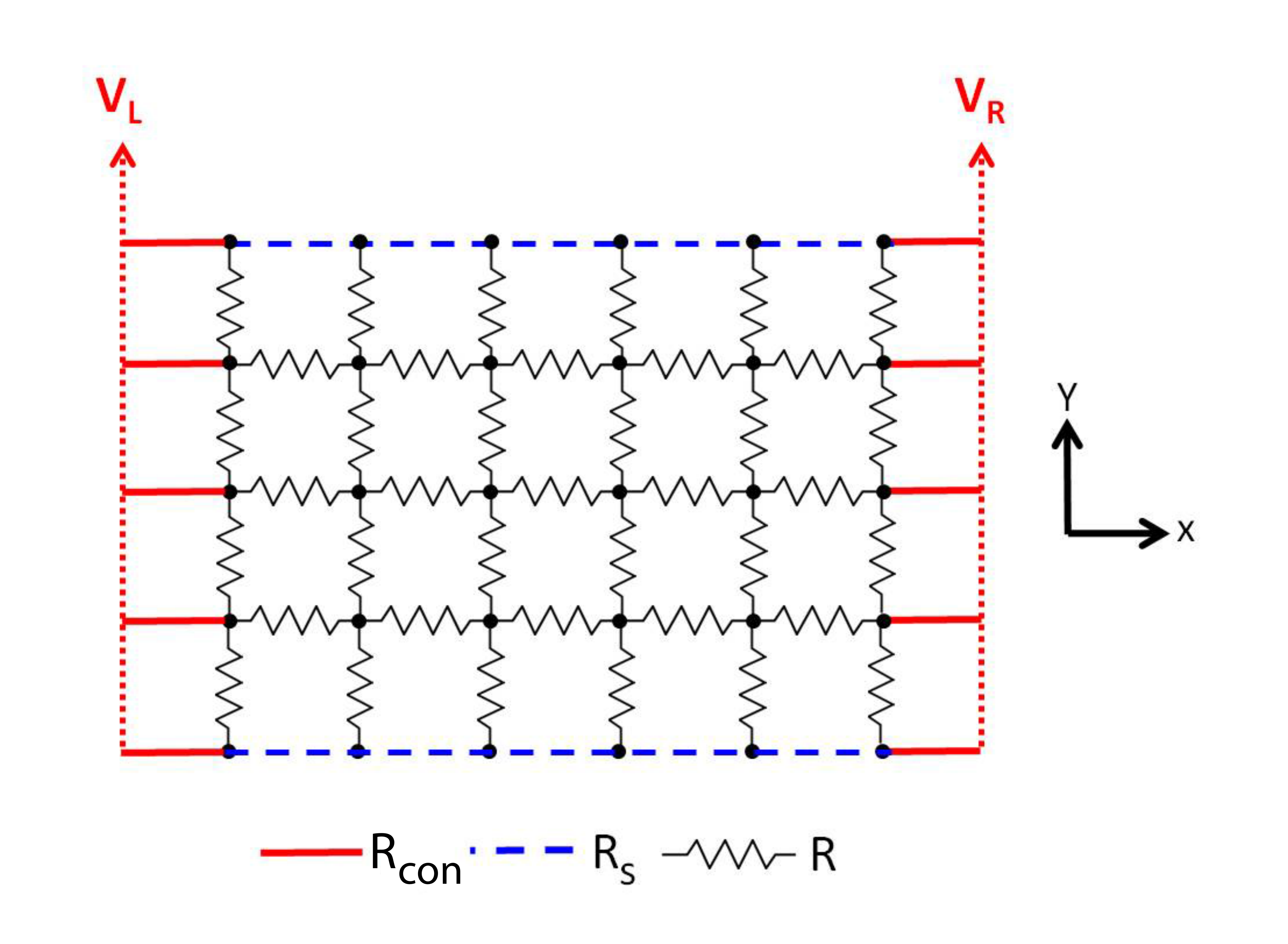}
\caption{Figure S12 $|$ A schematic of rectangular resistor network considered in this work. $N_x=6$ and $N_y=5$ were chosen for illustration purpose. In the numerical calculation, the boundary voltage $V_L>V_R$ are kept constant along the y direction.}
\end{center}
\end{figure}
The current $I_{0}$ is applied between the boundaries through the resistive contacts. Numerically this is implemented by setting the voltage at the left and right boundary sites, to $V_{L}$ and $V_{R}$, respectively, constant along the y direction. 
For every site $i$ the voltage $V_{i}$ satisfies Kirchoff\textquoteright{}s current law $\sum_{j}\frac{V_{i}-V_{j}}{R_{ij}}=0$, where $\{j\}$ are the neighbouring sites connected to site $i$. 
Thus, we have a system of linear equations for voltage $\{V_{i}\}$ which can be solved numerically as described elsewhere \cite{Kirkpatrick1973,Muniz2011}{}.
The current flow between two adjacent sites $i$ and $j$ can be determined as $I_{ij}=(V_{j}-V_{i})/R_{ij}$, where $R_{ij}=R$ for the bulk and $R_{ij}=R_{s}$ for the surface. 

Here, we are mainly interested in the current flow at the top surface.
Let $n_{x}^{(s)}=1,2,3$ ... $N_{x}$ and $V(n_{x}^{(s)})$ denote the coordinates and voltage of the surface sites. 
The surface current at site $n_{x}^{(s)}$ is determined as

\[
\Delta I_{s}(n_{x}^{(s)})=\frac{V(n_{x}^{(s)}+1)-V(n_{x}^{(s)})}{R_{s}}
\]which equals the current flowing from site $n_{x}^{(s)}$ to site $n_{x}^{(s)}+1$ in the limit $R_{s}\rightarrow0$.

In our calculation, we set $\sigma_{ab}/\sigma_{c}=10^{4}$ which is a reasonable value for LBCO-0.120 near the surface superconducting transition temperature (Fig. S7d), and $L_{x}:L_{y}=6:1$ due to the sample dimension a x c = 3.0 mm x 0.5 mm typically used in this work.
Based on these values, we derive the mesh aspect ratio

\[
N_{x}:N_{y}=L_{x}:L_{\tilde{y}}=L_{x}:\sqrt{\sigma_{ab}/\sigma_{c}}L_{y}\approx1:17
\]
In addition, we have $R_{con}\gtrsim R_{ab}$ where $R_{ab}=R\times N_{x}$ since the contact resistance is typically slightly higher than the sample resistance (see e.g. the resistance tomography results in Fig. S8a). 
This sets the resistor ratio $R_{s}:R:R_{con}\approx0:1:N_{x}$ for our numerical analysis.

Following the physical constraints described in the previous paragraph,we set $N_{y}=17N_{x}$, $R_{s}=0.0001$, $R=1$, $V_{L}=1$, and $V_{R}=0$, with $N_{x}$ and $R_{con}$ being the variable parameters in the calculations. 
Figure S13a shows the numerical results of $\Delta I_{s}$ for the case of $R_{con}/R_{ab}$=3 for different mesh sizes $N_{x}$=50, 100, 200.
The surface current $\Delta I_s$ exhibits approximately parabolic curve along the surface, with the peak value at the center, which is equal to 2.3\% of the total applied current $I_{0}$. 
It is important to note that the curve of $\Delta I$ is consistently reproduced for different values of $N_{x}$. This indicates that numerical error due to finite mesh size is insignificant in our calculation.

To examine how contact resistance affects the magnitude of the surface shunting current, we analyze for the case of $R_{con}/R_{ab}$=1, 3, and 10. Our results in Fig. S13b show that as $R_{con}/R_{ab}$ becomes smaller, the surface shunting current increases \textit{but} the bulk current remains finite and sizable. This reflects the fact that electrical contacts in our measurement prevents complete shorting of current to the superconducting surface. 

\begin{figure}[H]
\begin{center}
\includegraphics[scale=1,trim = 30 0 0 0]{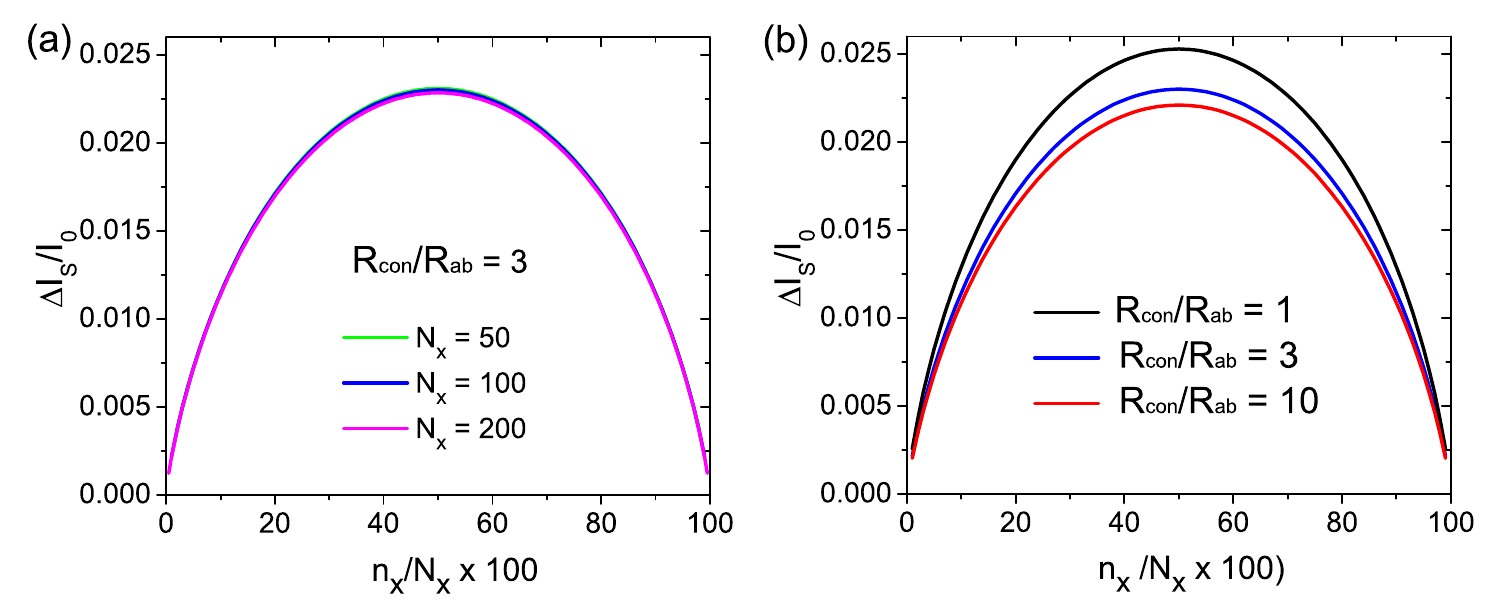}
\caption{Figure S13 $|$ Surface current $\Delta I_s$  as a function of site position $n_{x}^{(s)}$. a, For $R_{con}/R_{ab}=3$, $\Delta I_s$ is consistently reproduced over $N_x$=50, 100, 200 , with the peak value reaching 2.3 $\%$ of the applied current $I_0$. b, For $N_x=100$, $\Delta I_s/I_0$ is slightly suppressed on increasing $R_{con}$, resulting in a change of 0.35 $\%$ in the peak value between the cases of $R_{con}/R_{ab}=1$ and $R_{con}/R_{ab}=10$.}
\end{center}
\end{figure}

Figure S14 depicts the current flow in the bulk for illustration purpose. The local current $\Delta I$ is plotted as vector, with the x and y components determined as

\[
\Delta I_{x}(n_{x},n_{y})=\frac{V(n_{x}+1,n_{y})-V(n_{x},n_{y})}{R}
\]

\[
\Delta I_{y}(n_{x},n_{y})=\frac{V(n_{x},n_{y}+1)-V(n_{x},n_{y})}{R}
\]
where $(n_{x},n_{y})$ denote the site coordinate. We note that only regions close to the surface are affected by the surface. This shows that the superconducting surface is ineffective at shunting the bulk current, consistent with the results of $\Delta I_s$ where less than 2.5\% of the applied current $I_{0}$ actually reaches the surface.

\begin{figure}[H]
\begin{center}
\includegraphics[scale=0.1,trim = 30 0 0 0]{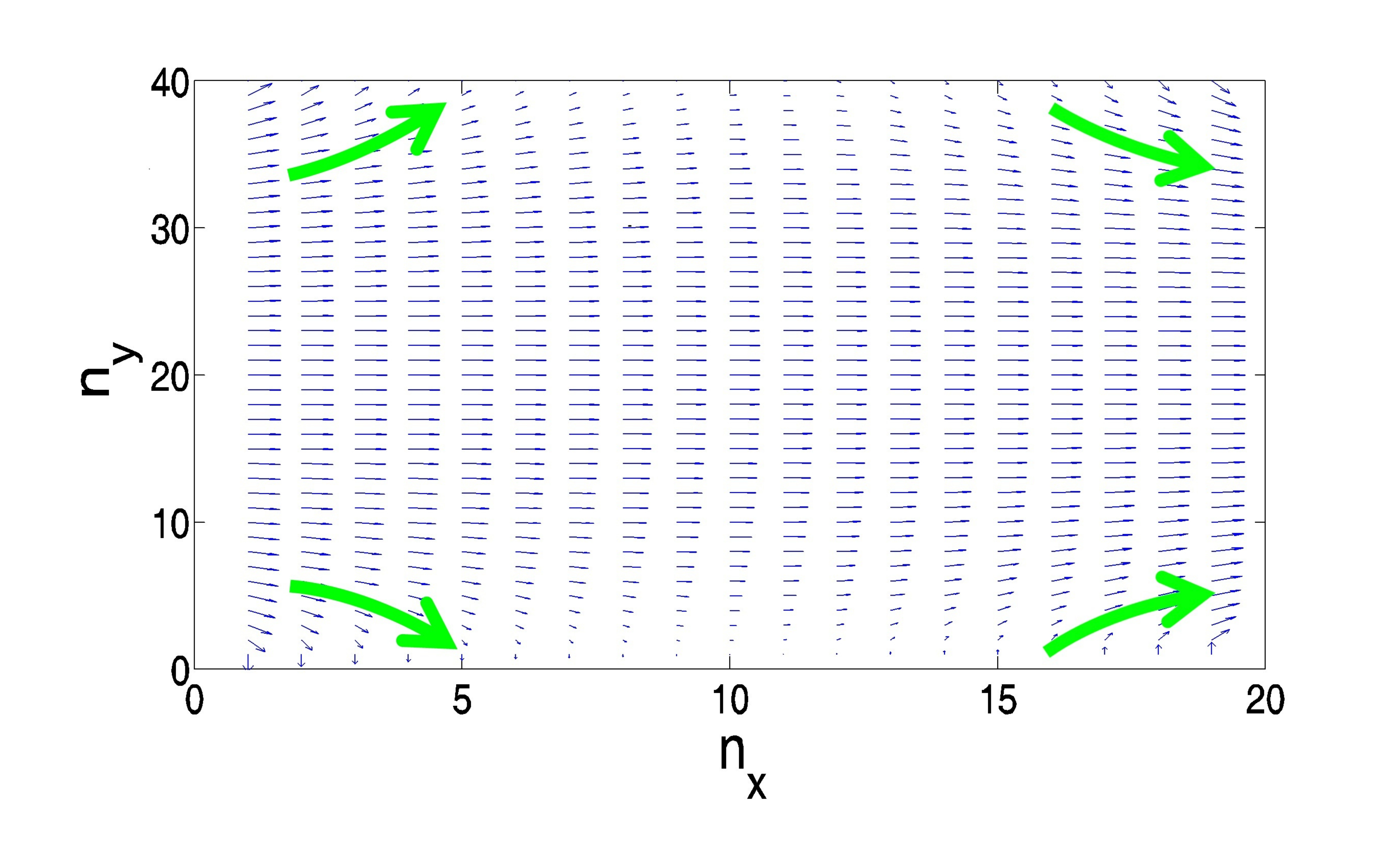}
\caption{Figure S14 $|$ The local current in the bulk $\Delta I$ plotted as a vector mesh. Here $N_x\times N_y=20\times 40$ is chosen for illustration purpose. The green arrows depict the local trends of the current flow in regions close to the surface.}
\end{center}
\end{figure}

\newpage
\section*{Mixture of Surface and Bulk Signals}
As discussed in section ``Resistor Network Model", the superconducting surface is ineffective in shunting the resistive bulk due to large c-axis resistance. As a result, with decreasing temperature the sample could evolve into an interesting state where the resistive bulk coexist with the superconducting surfaces.

In the measurements reported in the main part of the paper, the voltage contacts were carefully confined to either the top or the side of the sample. Meticulous care was taken to keep the voltage contacts a distance $\sim\mu$m from the sample edges to avoid pick-up of the counterpart signals. This ensured the surface and bulk were probed selectively enabling differentiation of the respective electronic transport. In the test case where the voltage contacts covered both the top and the side of sample, a combination of surface and bulk signals was measured. 

Figure S15a shows an example of resistance measurement where contributions from $\rho_s$ and $\rho_{ab}$ were observed, as indicated by the partial drop at $T_{cs}$ and the bulk superconducting transition at $T_c$. The measured resistance $R(T)$ can be reproduced by assuming a linear combination of bulk and surface in-plane resistivity, namely $R'(T)=A\rho_{ab}(T)+B\rho_s(T)$.
This excellent fitting explains the partial drop in $R(T)$ as a pickup of surface signal which vanishes at $T_{cs}=36$ K due to surface superconducting transition. For $T<T_{cs}$, $R(T)$ becomes purely bulk signal with excellent scaling to $\rho_{ab}(T)$.

The reason that superconducting surface does not completely short out the voltage measurement below $T_{cs}$ for such contact configuration is the presence of c-axis and contact resistances.
Figs S15b schematically illustrates the test case where voltage contacts cover both the top and the side of that sample.
The shunting effect by the superconducting surface is weakened by these resistive connections. For more detailed calculations, refer to the section ``Resistor Network Model"

\begin{figure}[H]
\begin{center}
\includegraphics[scale=0.35,trim = 30 0 0 0]{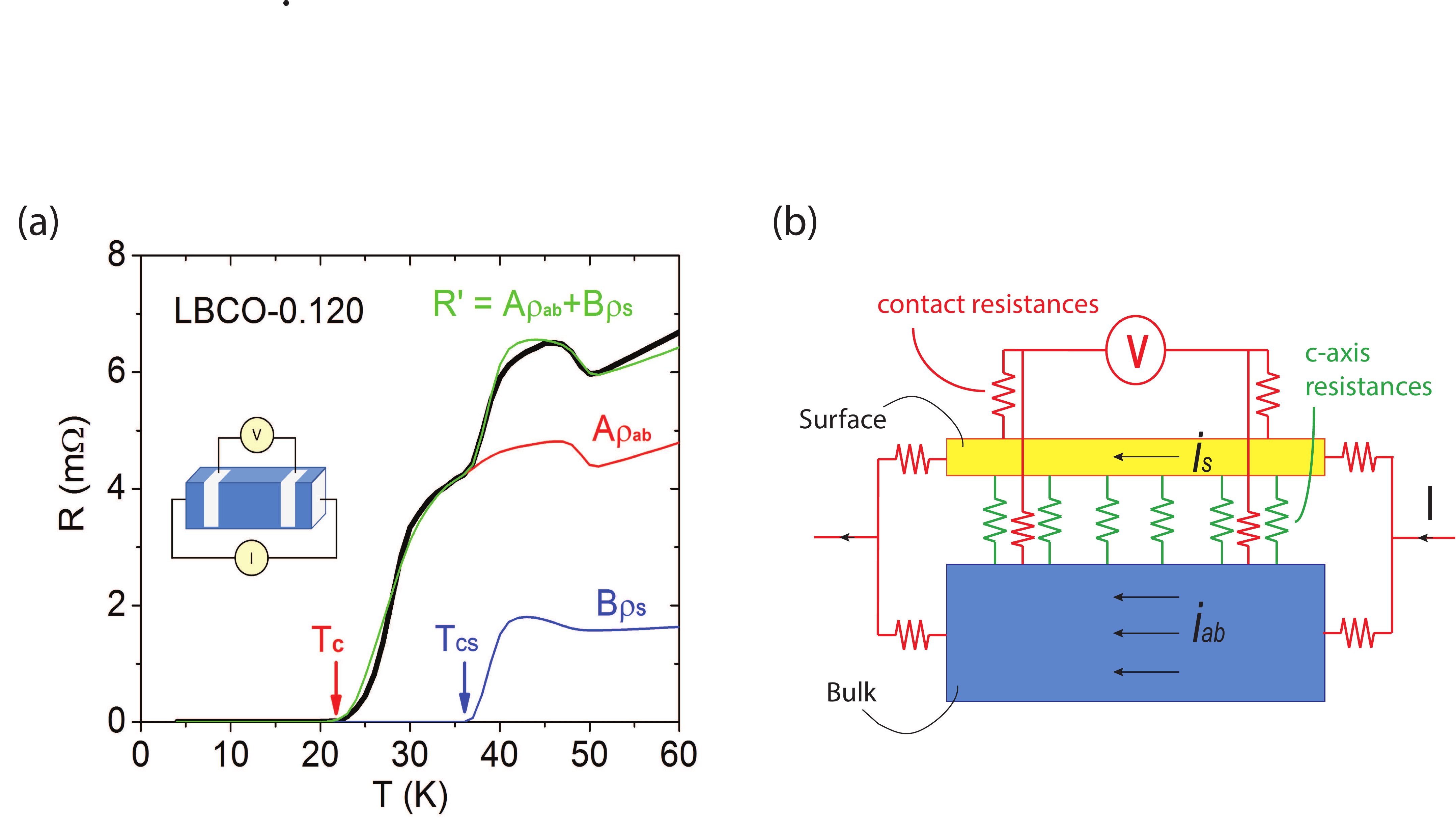}
\caption{Figure S15 $|$ Mixed contributions from \boldmath$\rho_s$ and \boldmath$\rho_{ab}$ in LBCO-0.120. a, In the test case where voltage contacts covered both the top and side of the sample (inset), the measured resistance (black line) showed a partial drop at $T_{cs}$ and eventually a complete one at $T_c$. It can be fit by $R'(T)=A\rho_{ab}(T)+B\rho_s(T)$ (green line) with $A=42.8$ and $B=7$, indicating contributions from both the surface and the bulk. b, Illustration of the resistive connections between the surface and the bulk, which weakens the current shunting to the surface.}
\end{center}
\end{figure}

\newpage
\section*{Ag Diffusion}
It is useful to understand how thick is the surface layer effectively probed by the contacts on the surface. 
At sufficiently high temperature, Ag may diffuse through multiple CuO$_2$ planes \cite{Dzhafarov96,Fang94}. In the case of Bi$_2$Sr$_2$CaCu$_2$O$_x$, diffusion in the c direction obeys the empirical function \cite{Fang94}

\begin{align*}
N(x,t)\approx \text{exp} \left(\frac{-x^2}{4Dt}\right)
\end{align*}
where $N$ is the normalized concentration of Ag in the sample, $x$ the distance from surface, $t$ the annealing time, and $D$ the c-axis diffusion coefficient. At $T=500$ $^{\circ}$C which is close to our annealing temperature 450 $^{\circ}$C, $D\approx3.9\times10^{-16}$ cm$^2$/s. Note that an additional contribution to $N(x,t)$ whose origin was not identified has been omitted in our analysis, since its magnitude is two orders smaller than the main component exp($-x^2/4Dt$).

To estimate the contact penetration in our samples, we assume the same diffusion behavior and coefficient of Bi$_2$Sr$_2$CaCu$_2$O$_x$. It is reasonable that contact resistivity increases as Ag concentration decreases into the sample. Consequently, not every Ag atom contributes to the electrical contacts. Here, we set $N(x_d,t)=0.01$ as the threshold criterion for electrical contact formation, which physically means the Ag concentration in the sample must not be less than 1\% of the actual level in the silver paste. This leads to
\begin{align*}
N(x_d,t)&\approx \text{exp} \left(\frac{-x^2_d}{4Dt}\right)=0.01\\
\implies x_d &=\sqrt{-4\text{ln}(0.01)Dt}
\end{align*}
For $D\approx 3.9\times 10^{-16}$ cm$^2$/s and $t=600$ s which is our annealing time, we obtain
\begin{align*}
x_d\approx 20 \text{ nm}
\end{align*}
which suggests that the electrical contacts penetrate approximately 20 nm into the sample. Remarkably, this number is close to our estimate of the superconducting surface thickness in the preceding section.

\bibliography{Bibliography/LBCO}